\documentclass{aa}  

\newcommand{\HCOp}{HCO$^+$}
\newcommand{\HnCOp}{H\element[][13]{C}O$^+$}
\newcommand{\DCOp}{DCO$^+$}
\newcommand{\Htp}{H$_3^+$}

\newcommand{\limescale}{0.7}
\newcommand{\thirdscale}{0.33}

\newcommand{\modr}{(R)} 
\newcommand{\moda}{(A)} 
\newcommand{\modb}{(B)} 

\usepackage{graphicx}
\usepackage{txfonts}

\usepackage{siunitx}
\usepackage{hyperref}
\hypersetup{
    colorlinks=true,
    linkcolor=blue,
    filecolor=magenta,      
    urlcolor=cyan,
    citecolor=blue,
}
\begin{document}


 \title{Using \HCOp{} isotopologues as tracers of gas depletion in protoplanetary disk gaps}

\titlerunning{Disk gaps traced by \HCOp{} isotopologues}

   \author{Grigorii V. Smirnov-Pinchukov
          \inst{1}
          \and
          Dmitry A. Semenov 
          \inst{1, 2}
          \and
          Vitaly V. Akimkin
          \inst{3}
          \and
          Thomas Henning
          \inst{1}
          }

   \institute{Max Planck Institute for Astronomy,
              K{\"o}nigstuhl 17, D-69117 Heidelberg, Germany\\
              \email{smirnov@mpia.de}
         \and
             Department of Chemistry, Ludwig Maximilian University, 
             Butenandtstr. 5-13, D-81377 Munich, Germany\\
        \and
            Institute of Astronomy, Russian Academy of Sciences, 
            Pyatnitskaya str. 48, Moscow, 119017, Russia
             }

   \date{Received \today; accepted}


  \abstract
   
  \abstract 
   {The widespread rings and gaps seen in the dust continuum in protoplanetary disks are sometimes accompanied by similar substructures seen in molecular line emission. One example is the outer gap at $\sim 100$\,au in AS~209, which shows that the \HnCOp{} and C$^{18}$O emission intensities decrease along with the continuum in the gap, while the \DCOp{} emission increases inside the gap.}
   {We aim to study the behavior of \DCOp{}/\HnCOp{} and \DCOp{}/\HCOp{} ratios in protoplanetary disk gaps assuming the two scenarios: \moda{} the gas depletion follows the dust depletion and \modb{} only the dust is depleted.}
   {We first modeled the physical disk structure using the thermo-chemical model ANDES. This 1+1D steady-state disk model calculates the thermal balance of gas and dust and includes the far ultraviolet (FUV), X-rays, cosmic rays, and other ionization sources together with the reduced chemical network for molecular coolants. Afterward, this physical structure was adopted for calculations of molecular abundances with the extended gas-grain chemical network with deuterium fractionation. Ideal synthetic spectra and 0th-moment maps were produced with the LIne Modeling Engine (LIME).}
   {We are able to qualitatively reproduce the increase in the \DCOp{} intensity and the decrease in the \HnCOp{} and C$^{18}$O intensities inside the disk gap, which is qualitatively similar to what is observed in the outer AS~209 gap. The corresponding disk model \moda{} assumes that both the gas and dust are depleted in the gap. The model \modb{} with the gas-rich gap, where only the dust is depleted, produces emission that is too bright in all \HCOp{} isotopologues and C$^{18}$O.}
   {The \DCOp{}/\HnCOp{} line ratio can be used to probe gas depletion in dust continuum gaps outside of the CO snow line. The \DCOp{}/C$^{18}$O line ratio shows a similar, albeit weaker, effect; however, these species can be observed simultaneously with a single ALMA or NOEMA setup.}


   \maketitle
%


\section{Introduction}
\label{txt:introduction}

Protoplanetary disks are believed to be the birthplaces of planetary systems. While more and more theoretical studies of the planet formation in disks appear in the literature, they need observational {constraints regarding} physical conditions and chemical composition in various disks and disk locations. Now, in the Atacama Large (sub)Millimeter Array (ALMA),  
NOrthern Extended Millimeter Array (NOEMA), 
 Very Large Telescope / Spectro-Polarimetric High-contrast Exoplanet REsearch (VLT/SPHERE), and  Gemini Planet Imager (GPI)
era, the spatially resolved observations of dust and gas in protoplanetary disks in nearby star-forming regions have become a routine. High-resolution studies in dust continuum and scattered light with an angular resolution up to $0.025''-0.04''$ reveal complex substructures with multiple gaps, inner holes, rings, and spirals \citep[DSHARP:][]{DSHARP1, DSHARP2,Long_2019_Taurus_survey,Huang_ea20a,2017ApJ...837..132V}.
The high-resolution observations in the CO isotopologues and other molecules also show the presence of substructures in the disk gas \citep[see, e.g.,][]{2016PhRvL.117y1101I, 2017A&A...600A..72F,Teague_ea_2017,Huang_ea_2018_TWHya}.  Often {disk} substructures that are {detected} in the scattered light at near-infrared wavelengths do not coincide with the substructures visible in the {submillimeter} dust continuum and gas emission lines \citep{2019A&A...625A.118K}. 

One of the peculiar disks showing very wide gaps is AS~209. \citet{Huang2017} have studied continuum and line emission in this disk at the $\sim 0.4''$ resolution with ALMA. They have found a notable difference between \HnCOp{} and \DCOp{} emission radial profiles, with the \DCOp{} emission peak extending farther out up to radii of $\lesssim 80-100$~au. The follow-up higher resolution observations have shown that the AS~209 disk has a dust gap at around 55--120~au \citep{2018A&A...610A..24F,Favre_ea_2019_AS209}. At this gap location, the optically thin C$^{18}$O J=2--1 emission decreases slightly in intensity, while optically thick $^{12}$CO J=2--1 does not show any intensity decreases, and the \DCOp{} J=3--2 line reaches its peak intensity. Inspired by these puzzling observational findings, we decided to study whether a combination of optically thin and thick \HCOp{} isotopologues could be used as a probe for gas depletion in disk gaps that have been detected before in dust continuum or CO isotopologues.

There are several major factors that can cause molecular line emission in disks to show  gap-like or ring-like substructures. Firstly, a circular gap or a ring in the line emission could be driven by a local, rather azimuthally symmetric deviation in the physical structure of the disk, either in density or temperature \citep[e.g., due to planet-disk interactions, snowlines, a change in dust properties, etc.][]{LP_93,Guzman_ea18a,Huang_ea20a}. In the case of a surface density drop, the total molecular column density could become smaller or higher (e.g., for ions), which would lead to weaker or stronger line emission, as long as the line remains optically thin. Similarly, a local increase or decrease in the gas's kinetic temperature or background radiation affect molecular line excitation and hence emission intensity, assuming that the line is thermalized. 

{Secondly, molecular emission could show rings or gaps caused primarily by various chemical effects. The most prominent ones are ultraviolet (UV)-driven production or (selective) photodissociation in the disk atmosphere, C/O variations, and the freeze-out of molecules onto the dust grain surfaces in the disk midplane
\citep[see, e.g.,][]{Dutrey_ea17,Teague_ea_2017,2018A&A...609A..93C,Miotello_ea18a,Miotello_ea19a,vTerwisga_ea19a,Garufi_ea20a}. For example, c-C$_3$H$_2$, CCH, CN, H$_2$CO, and CS emission often shows a ring-like appearance in disks. Whereas for hydrocarbons and CN emission, the C/O variations and high-energy UV or X-ray irradiation are the most important factors; for CS and H$_2$CO emission, the surface chemistry processes and freeze-out play the major role
\citep{Bergin_ea16,2018A&A...609A..93C,Miotello_ea19a,vTerwisga_ea19a}.
The ring-like appearance of N$_2$D$^+$ and DCO$^+$ emission in disks is mainly due to low-temperature deuterium fractionation when gaseous CO freezes out in the disk midplane \citep{Qi_ea13a,Ceccarelli_ea14,Huang_Oberg15,Salinas_2018_DCOp,Garufi_ea20a}. In addition, temporal variations in physical conditions can also affect chemical abundances, for example, via X-ray flares or episodic accretion outbursts \citep[e.g.,][]{Cleeves_ea17,2018ApJ...866...46M,Lee_ea19a}.}

{Finally, a third relevant effect for molecular line excitation and emission intensity is the departure from local thermal equilibrium (LTE). It depends on the molecular properties and the location of the emitting molecular layer in the disk \citep{2007ApJ...669.1262P}. The most notable examples are the CN radical emitting from upper disk layer, where excitation is dominated by radiative processes,
and CH$_3$OH, which have a complex energy level structure with torsional and coupled energy levels \citep[e.g.,][]{2016MNRAS.460.2648P,2018A&A...609A..93C,Teague_Loomis20}. Another effect that may affect the line excitation in disks locally is the localized deviation in the physical structure of the disk, which does not affect the global structure of the disk otherwise \citep[e.g., local heating in a circumplanetary disk][]{Cleeves_ea15}. This shows the
complex interplay between the physical structure, chemistry, and excitation
conditions in defining the strength of line emission and its spatial distribution in a disk.}

The paper is structured as follows. In the Section~\ref{txt:method}, we present the tools we use and the assumptions we make. In Section~\ref{txt:results} we present the results for each model (\ref{txt:res-r}-\ref{txt:res-d}). In Section~\ref{txt:discussions} we discuss the model uncertainties (\ref{txt:uncertainties}) and the comparison with previous observational gas gap studies (\ref{txt:comparison}). We provide the final conclusion in Section~\ref{txt:conclusion}.

\section{Methods and models}
\label{txt:method}

{In this section we list all of the steps involved in our modeling. Firstly, we describe ANDES, the thermo-chemical code we used to simulate 1+1D disk gas and dust thermal structure (Subsection~\ref{txt:thermo-chemical-modeling}). Secondly, we explain how we combined the simulated vertical columns into three 2D disk models, with one reference model and two models with and without gas depletion in the dust gap. We recalculated the radial ionization and photodissociation self-shielding factors for those 2D physical structures (\ref{txt:post-processing}). Thirdly, we briefly overview the chemical model ALCHEMIC with the deuterated network, which we used to calculate the chemical abundances (\ref{txt:chemistry}). Finally, we describe how we computed the synthetic molecular line emission maps with LIME (\ref{txt:lime}).} 

\subsection{Thermo-chemical modeling}
\label{txt:thermo-chemical-modeling}
The standard practice of chemical modeling of protoplanetary disks includes some density structure setup, dust radiative transfer simulations for the thermal structure of the dust, gas thermal balance, and chemical kinetics calculations using the resulting structure (Dust And LInes (DALI) \citep{Bruderer_ea_2009a, Bruderer_ea_2014_DALI2}, PRotoplanetary DIsk MOdel (ProDiMo) \citep{Woitke_Kamp_Thi_2009_ProDiMo}, ANDES \citep{Akimkin_ea_2013}, also \citep{2004ApJ...613..424G, Cleeves_ea_2013,2014ApJ...792....2D,2015A&A...582A..88W,Salinas_2018_DCOp}).
The ANDES model allows one to simulate the chemical evolution of a protoplanetary disk with a detailed treatment of gas and dust thermal balance. Its original version \citep{Akimkin_ea_2013} is based on a 1+1D approach, where the vertical disk structure at each radius is simulated independently. In the upper, low-density regions of the protoplanetary disk, the gas temperature is not coupled to the dust temperature. The gas density becomes too low for thermal accommodation with the dust, and the gas usually becomes hotter than the dust as it cools itself rather inefficiently \citep{Roellig_gas_thermal_balance}.
The gas density, temperature, and chemical composition depend on each other, so iterative modeling of the vertical disk structure is performed. After the initial chemical assumptions are made, the thermal balance for this predefined chemical structure is calculated. The chemical kinetics calculations are iterated with thermal balance until convergence in further iterations. In the utilized version of the ANDES code, the dust temperature step is recalculated again for the hydrostatic solution that fits the vertical temperature structure{, assuming well-mixed small dust particles and, therefore, a constant dust-to-gas mass ratio}.

The recent version of the {ANDES$_{\rm 2D}$} code is based on a fully 2D approach and it simultaneously iterates the vertical disk structure {and the chemical evolution}, which allows {one} to accurately simulate some time-dependent effects {such as} luminosity outbursts in the FU~Ori-type systems \citep{2017ApJ...849..130M, 2018ApJ...866...46M}. However, this recent version does not feature the detailed gas thermal balance. In this study, we use the original version of  ANDES  with the reduced chemical network for molecular coolants and modifications to the modeling of the X-ray and cosmic ray ionization rates. The reduced chemical network consists of 63 species and 480 reactions to accurately model the abundances of C, C$^+$, H$_2$O, CO, CO$_2$, OH, H$_2$, and other simple species. It has been obtained from the full network (described in Subsection~\ref{txt:chemistry}) by using the iterative reaction-based reduction method followed by manual tuning \citep{Wiebe_ea03a,Semenov_ea04}.

Using the ANDES model, we first computed a set of disk thermo-chemical models to obtain 1+1D temperature and density structures. We assumed the stellar and disk parameters, as listed in Table \ref{tab:andesinp}.  We selected a T~Tauri star with a mass of $M=0.5\ M_\sun$, and we derived its effective temperature $T_{\rm eff}=3820\ \rm K$ and radius $R=1.47\ R_\sun$ from the evolutionary track model assuming an age of 2 Myr \citep[][private communication]{Yorke_Bodenheimer_2008_track}. We did not aim to qualitatively model the K5e-type AS~209 star, rather we {modeled} a generic T~Tauri protoplanetary disk. For the UV excess, we assumed the blackbody emission with $T_{\rm eff} = 20000$ K, normalized to the total flux of blackbody radiation with $T_{\rm eff} = 4000$ K. We did not take the evolution of the stellar radius and effective temperature into account. For the dust radiative transfer and surface chemistry, we used non-evolving single-size silicate grains with a size of $10^{-5}$\,cm. We set the dust-to-gas mass ratio to 0.01 in our reference disk model.

The disk surface density was parameterized using the following power-law distribution, without taking tapering at outer radii into account: 
\begin{equation}
\Sigma(r) = \Sigma_{1 \mathrm{au}} \left( \frac{r}{1\ \mathrm{au}}\right)^{-1}
\label{eq:coldens}
.\end{equation}

{This approach leads to a denser outer disk and affects the integral disk properties. On the other hand, it does not affect the effect of the local substructure much.
We also separately computed radii between 80 and 250\,au with a 10 times lower surface density, and with a 10 times lower dust-to-gas ratio for the gap models, which is described in the next section.
} The summary of the adopted stellar and disk parameters is presented in Table~\ref{tab:andesinp}.

   \begin{table}
       
      \caption[]{ANDES input parameters.}
         \label{tab:andesinp}
         \centering 
        \begin{tabular*}{\linewidth}{@{\extracolsep{\fill}}ll}

            \hline \hline
            \noalign{\smallskip}
            Parameter      &  Value \\
            \noalign{\smallskip}
            \hline
            \noalign{\smallskip}
            $M_{\rm star}$ & $0.5\ M_\sun$ \\
            $T_{\rm star}$ & 3820 K   \\
            $R_{\rm star}$ & $1.47\ R_\sun$ \\
            UV excess & $B_\nu(T_{\rm eff}= 20000\ \mathrm{K}) $\\
            Accretion rate & \num{3.6e-9} $ M_\sun \rm{yr}^{-1}$ \\
            \noalign{\smallskip}
            \hline
            \noalign{\smallskip}
            Dust opacity & \citet{1984ApJ...285...89D} \\
            Dust grain size & $10^{-5}$\ \text{cm} \\
            Dust-to-gas mass ratio & 0.01, 0.001 \\
            Grazing angle & 0.05 \\
            Gas column density at 1\ \text{au} & 10, 100 g cm$^{-2}$\\
            Density power-law slope & $-1$\\
            \noalign{\smallskip}
            \hline
        \end{tabular*}

   \end{table}

\subsection{ANDES post-processing}
\label{txt:post-processing}
\subsubsection{Models with gaps}
\label{txt:gaps}
\begin{figure*}
\centering
     Model \modr{}\hspace{0.25\linewidth} Model \moda{}\hspace{0.25\linewidth} Model \modb{}
\includegraphics[width=\thirdscale\linewidth,clip,page=70]{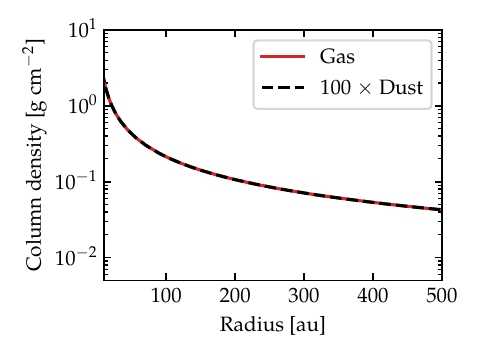}
\includegraphics[width=\thirdscale\linewidth,clip,page=70]{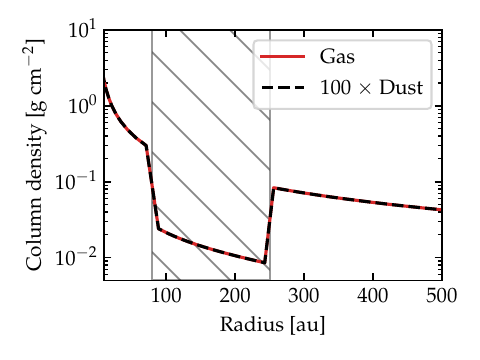}
\includegraphics[width=\thirdscale\linewidth,clip,page=70]{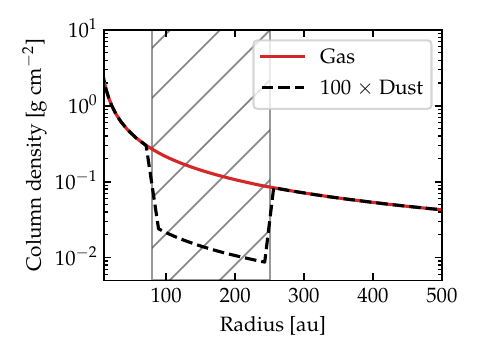}
\includegraphics[width=\thirdscale\linewidth,clip,page=48]{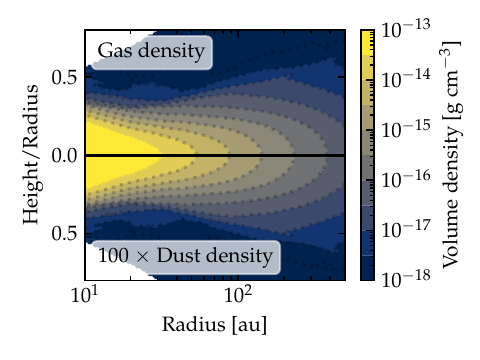}
\includegraphics[width=\thirdscale\linewidth,clip,page=48]{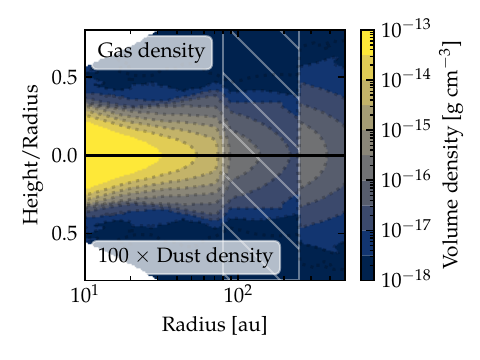}
\includegraphics[width=\thirdscale\linewidth,clip,page=48]{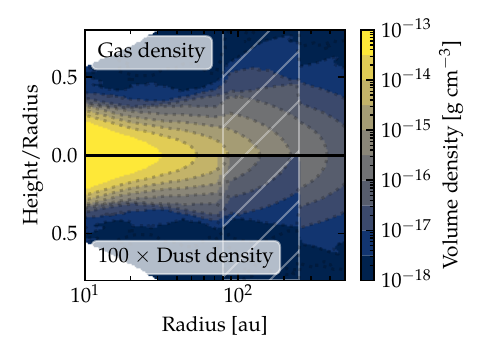}
\includegraphics[width=\thirdscale\linewidth,clip,page=43]{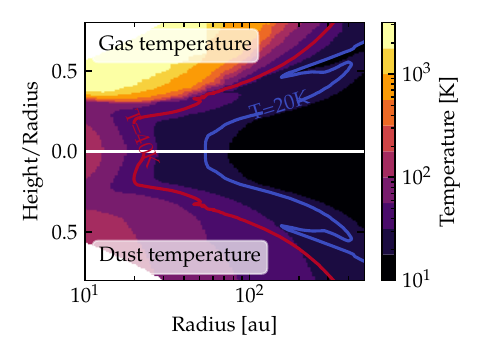}
\includegraphics[width=\thirdscale\linewidth,clip,page=43]{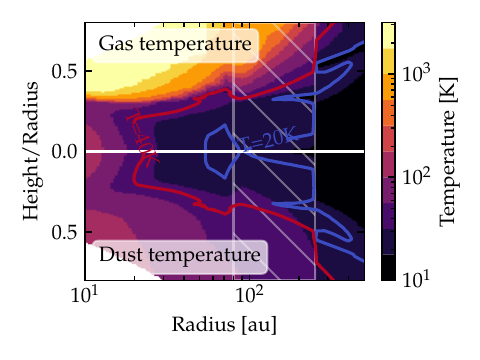}
\includegraphics[width=\thirdscale\linewidth,clip,page=43]{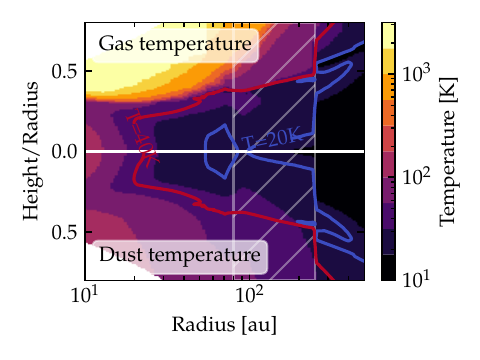}
\caption{Physical structure of the three protoplanetary disk models. The left panels are for the reference model \modr{}, the middle panels are for the gas-poor gap model \moda{}, and the right panels are for the gas-rich gap model \modb. {In the upper row, the column densities of gas and dust are presented. The dust column density is multiplied by 100 for comparison purposes with the gas and the typical dust-to-gas mass ratio of 0.01}. In the {second} row, the upper half of each panel is the gas density, and the bottom half is the dust density, multiplied by 100 for comparison. In the bottom row, the upper half of each panel is the gas temperature, and the bottom half is the dust temperature. The location of the gap is marked by a hatched rectangle. The 20 and 40\,K gas temperature isotherms are shown as blue and red contour lines, respectively.
}
\label{thermal_structure}
\end{figure*}

   \begin{table*}[ht]

      \caption[]{Properties of the disk models.}
         \label{tab:models}
         \centering 
        \begin{tabular*}{\linewidth}{@{\extracolsep{\fill}}lllll}

            \hline \hline
            \noalign{\smallskip}
            Model      &  Gas gap & Gas column density & Dust gap & Dust-to-gas mass ratio  \\
            & [au] & (inside the gap) &  [au]  &  (inside the gap) \\
            \noalign{\smallskip}
            \hline
            \noalign{\smallskip}
            \modr{} Reference disk & no & $100\ \text{g cm}^{-2} \left(\frac{r}{ 1\, \text{au}}\right)^{-1}$&  no& 0.01 \\
            \moda{} Gas- and dust-poor gap & $80-250$ & $0.1\ \times\ $ (Reference) &$80-250$& 0.01 \\
            \modb{} Gas-rich, dust-poor gap & no & (Reference)  & $80-250$& 0.001 \\
            \noalign{\smallskip}
            \hline
        \end{tabular*}

   \end{table*}

After that, we combined the vertical structures computed by ANDES for each disk radii into the three protoplanetary disk models (see Table~\ref{tab:models}). The first {model}, the so-called reference model, hereafter referred to as \modr{}, is the disk model without any dust or gas gaps. For the other two disk models, we used {the} separately computed vertical structures for the radii between 80 and 250\,au, and those from the reference model \modr{} outside of this gap. For the second model, the so-called gas-poor gap model, hereafter referred to as \moda{}, we lowered gas and dust surface densities by a factor of 10 between 80 and 250\,au, such that the dust-to-gas mass ratio remains the same inside the gap as in the reference model \modr{}. For the third {model}, the so-called gas-rich gap model, hereafter referred to as \modb{}, we only lowered the dust surface density by a factor of 10 between 80 and 250\,au, such that the gas surface density remains the same inside the gap as in the reference model. This setup leads to a much lower dust-to-gas mass ratio of 0.001 inside the gap in model \modb{}. Our 1+1D approach does not include the interaction between different radii, which are treated independently; furthermore, the physical conditions and abundances at the gap's edges may not be completely feasible. We present the resulting density and thermal disk structures in Fig.~\ref{thermal_structure}.
The width of the gap used in our modeling is larger than a typical gap width of $\sim 1-20$~au observed in the dust continuum, but it is similar to the wide gap found in the AS~209 disk. The effects of narrower gaps of a similar depth located anywhere between 80 and 250\,au on the disk physics and chemistry would be similar {because of the 1+1D nature of our modeling tools}. 

\subsubsection{X-rays, cosmic rays, and stellar wind}
\label{txt:ionization}

{Other than} cosmic rays, stellar X-ray radiation is one of the primary ionization mechanisms in protoplanetary disks. X-rays are produced by the accreting gas and its interaction with magnetic fields (see \citet{Rab_ea_2018} and references therein). In addition to X-rays, it has recently been recognized that the high-energy stellar energetic particles (SEP) could also play an important role in the disk ionization \citep{Rab2017}.  We added the SEP ionization using the active T~Tauri model of \citet{Rab2017}. While far ultraviolet (FUV) radiation dominates ionization in the disk atmosphere \citep{Roellig_gas_thermal_balance}, both X-rays and SEP are the primary ionization sources in the intermediate layers of the disk \citep{Semenov_ea04}. In the midplane, the ionization is dominated by the cosmic rays or the decay of short-lived radionuclides (SLRs). 

The X-ray ionization rate at a specific disk location primarily depends on the X-ray flux and the hardness of the X-ray energy spectrum. To compute the X-ray flux at a given point in the disk, one should perform X-ray radiative transfer {modeling}. The addition of the Compton scattering is necessary for more realistic calculations of the X-ray ionization rates in the disk. However, the Monte-Carlo X-ray radiative transfer is computationally expensive.

Instead, a more conventional approach is the computation of the X-ray ray tracing on  simplified 1D or 2D geometries and the further usage of parametrization to approximate the disk structure. In recent versions of the ProDiMo code, the X-ray radiative transfer was performed for every disk structure \citep{Rab_ea_2018}.  In the DALI code, the Compton scattering is neglected, and the one-dimension ray tracing from a point-like stellar source is assumed \citep{Bruderer_ea_2009a}. 

The parametrization of the disk's X-ray irradiation given in \citet{Bai_Goodman_2009} is based on the radiative transfer model of two bremsstrahlung-emitting coronal rings located at the height of $10\ R_\sun$ from the rotation axis and a similar distance above and below the disk midplane. This model also takes the Compton scattering and photoionization absorption into account. \citet{Bai_Goodman_2009} followed the approach of \citet{Igea_Glassgold_1999}, who studied the X-ray ionization of the inner (on the order of a few au) regions of protoplanetary disks. \citet{Igea_Glassgold_1999} found that the ionization structure can be parameterized as a function of only vertical column density for a given radius by scaling down an unattenuated X-ray flux at the disk atmosphere.
We adopted the simplified approach of \citet{Bai_Goodman_2009} in our ANDES 1+1D iterative modeling of the disk structure. After the iterative calculations of the disk were finished, we utilized the more rigorous method of  \citet{Bruderer_ea_2009a} and recalculated the disk ionization structure, assuming a single-point X-ray source associated with the central star. The X-ray shielding of the gap by the inner disk regions was computed by the 2D ionizing radiation ray tracing without scattering.

Finally, Galactic cosmic rays (CRs) dominate the ionization rate in the disk midplanes  when the local gas surface density does not exceed about 100~g\,cm$^{-2}$
or when the stellar wind and magnetic mirroring are not that strong \citep{DS_94,Gammie_96,Cleeves_ea_2013,Rab2017, 2018A&A...614A.111P}. Otherwise, in the densest disk regions shielded from CRs, the decay of SLRs plays a major role. We used the unattenuated CR ionization rate of $1.3 \times 10^{-17}$~s$^{-1}$ and
the SLR ionization rate of $6.5 \times 10^{-19}$~s$^{-1}$ \citep{Semenov_ea04}. In the studied regions of our adopted disk models, neither the gas surface density nor the stellar wind is strong enough to efficiently block the CRs, so the impact of SLRs on the disk midplane ionization is negligible.

Last but not least, for each iterative modeling step, the radial CO and H$_2$ column densities were calculated. They were used to compute the UV shielding factors, 
using an interpolation table from \citet{1996A&A...311..690L} \citep[see also][]{Semenov_Wiebe11, Visser_ea09b}.

\subsection{Chemical model}
\label{txt:chemistry}


Next, we used the computed physical structures of the disk and performed time-dependent chemical  modeling as post-processing with the ALCHEMIC code \citep{Semenov_ea_2010}. The gas-grain network is based on the osu.2007 rate file with updates as of June 2013 from the \textsc{kida} database\footnote{\tt http://kida.obs.u-bordeaux1.fr} (see \citet{KIDA}). The network is supplied with a set of approximately $1\,000$ reactions with high-temperature barriers from \citet{Harada_ea10} and \citet{Harada_ea12}.
This network was extended by using a statistical branching approach with a set of deuterium fractionation reactions and it includes up to triply-deuterated species 
\citep[see][]{Albertsson_ea13,Albertsson_ea14a}. Primal isotope exchange reactions for H$_3^+$, CH$_3^+$, and C$_{2}$H$_{2}^+$ from \citet{2000A&A...361..388R}, \citet{GHR_02}, \citet{2004A&A...424..905R}, and \citet{2005A&A...438..585R} were included. Ortho and para forms of H$_2$, H$_2^+$, and H$_3^+$ isotopologues were considered. The 
corresponding nuclear spin-state exchange processes were added from  experimental and theoretical studies  \citep{1990JChPh..92.2377G,2004A&A...427..887F,2004A&A...418.1035W,2006A&A...449..621F,2009A&A...494..623P, 2009JChPh.130p4302H,2011PhRvL.107b3201H,2013A&A...554A..92S}.
 

The assumed grains are spherical nonporous silicate particles with the material density of $3$~g\,cm$^{-3}$ and a radius of $0.1\,\mu$m. Each grain provides around $1.88\times10^6$ surface sites \citep[][]{Bihamea01}. 
The gas-grain interactions include the freeze-out of neutral species and electrons to dust grains with a 100\% sticking probability. 
In addition, the dissociative recombination and radiative neutralization 
of molecular ions on charged grains and grain recharging are taken into account.
As desorption mechanisms for ices, we considered thermal, CR-, UV-driven, and reactive desorption processes. 
An FUV photodesorption 
yield of $1\times 10^{-5}$ was adopted. In addition, FUV-driven photodissociation reactions inside ice mantles were implemented as in \citet{Garrod_Herbst06} and \citet{Semenov_Wiebe11}. 
Surface recombination is assumed to proceed via the classical Langmuir-Hinshelwood mechanism \citep[e.g.,][]{HHL92}. The ratio between 
diffusion and desorption energies of surface reactants is assumed to be 0.4. Hydrogen tunneling through rectangular reaction barriers with a thickness of $1\, \AA$ was computed using Eq.~(6) from \citep{HHL92}. For each act of surface recombination, the efficiency of reactive desorption of the reaction products directly into the gas phase was assumed to be 1\% \citep{2007A&A...467.1103G,2013ApJ...769...34V}.

Our chemical network consists of 1268 species made of 13 elements and 38812 reactions.  As initial abundances, we used the so-called low metals set of mainly atomic abundances (except H$_2$ and HD) from \citep{Lee_ea98}. Table~\ref{tab:initial_abundances} summarizes the initial relative abundances (wrt the total amount of hydrogen nuclei). The ALCHEMIC code and the chemical network were used to model the chemical evolution of the disk over an evolutionary time span of 1~Myr
for all three physical structures that were considered.

{With our large gas-grain chemistry network with deuterium fractionation, it was not practical to add another full set of $^{13}$C and $^{18}$O reactions. That is why we opted for simple scaling relations when deriving abundances of $^{13}$C and $^{18}$O-bearing isotopologues of CO and \HCOp{}. We assumed that the \HnCOp{} abundances are equal to 1\% of the \HCOp{} abundances, while C$^{18}$O abundances are equal to 0.2\% of the $^{12}$CO abundances \citep{Wilson_Rood94}. This simplistic approach does not take selective photodissociation of CO isotopologues and isotopic $^{13}$C-exchange reactions into account, which may lead to uncertainties in the adopted \HnCOp{} and C$^{18}$O abundances by a factor of 3--5 \citep[e.g., ][]{Visser_ea09b,Miotello_ea14,Miotello_ea18a}. Since we mainly focus on analyzing the behavior of abundance and line ratios, rather than absolute values, such a discrepancy should not affect the major underlying trends.}

\begin{figure*}
\centering

      Model \modr{}\hspace{0.25\linewidth} Model \moda{}\hspace{0.25\linewidth} Model \modb{}
\includegraphics[width=\thirdscale\linewidth,clip,page=50]{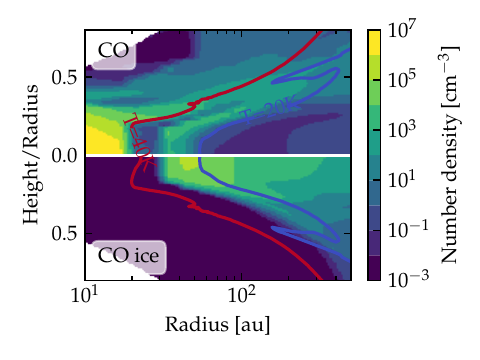}
\includegraphics[width=\thirdscale\linewidth,clip,page=50]{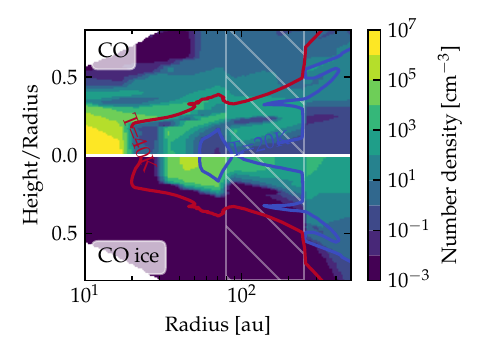}
\includegraphics[width=\thirdscale\linewidth,clip,page=50]{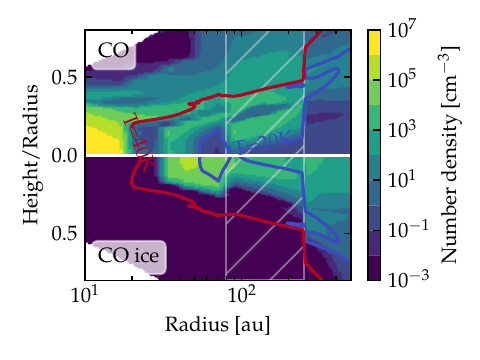}

\includegraphics[width=\thirdscale\linewidth,clip,page=67]{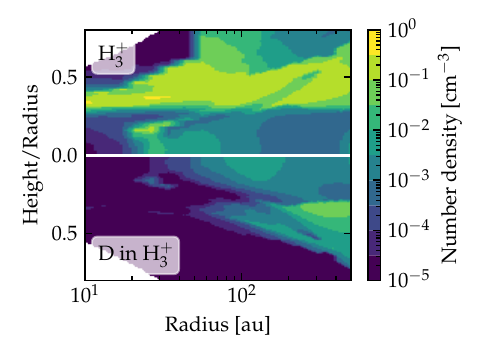}
\includegraphics[width=\thirdscale\linewidth,clip,page=67]{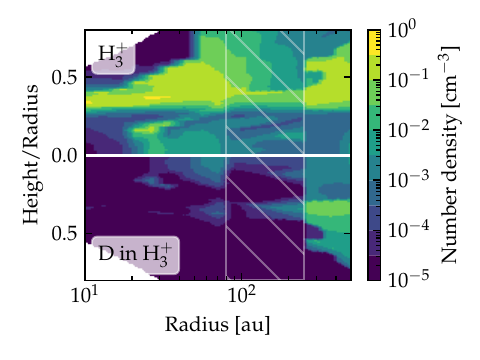}
\includegraphics[width=\thirdscale\linewidth,clip,page=67]{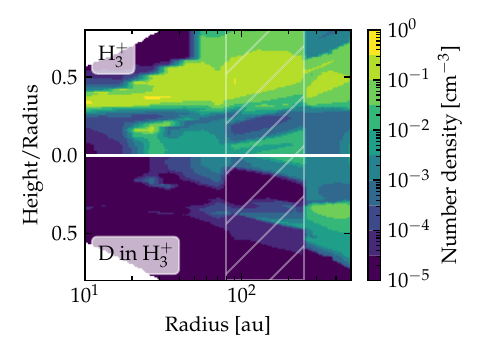}

\includegraphics[width=\thirdscale\linewidth,clip,page=42]{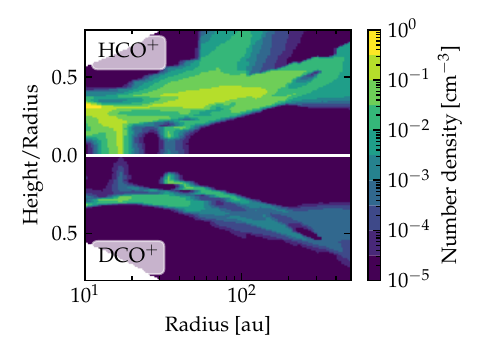}
\includegraphics[width=\thirdscale\linewidth,clip,page=42]{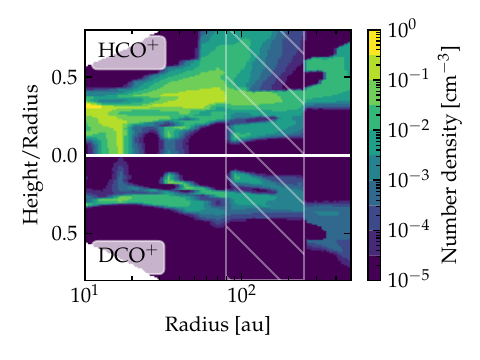}
\includegraphics[width=\thirdscale\linewidth,clip,page=42]{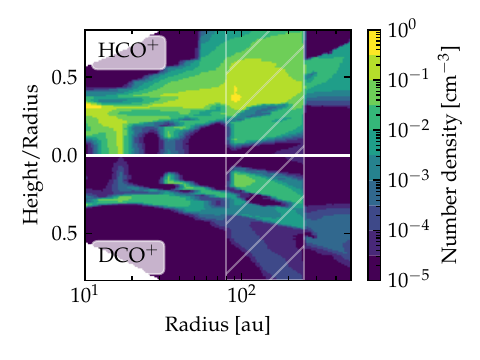}

\caption{Chemical structures of the three protoplanetary disk models. The left panels are for the reference model \modr{}, the middle panels are for the gas-poor gap model \moda{}, and the right panels are for the gas-rich gap model \modb. Different species are shown in the top and bottom halves of each panel. D in \Htp{} is the total number of D atoms in \Htp{} isotopologues. The location of the gap is shown as a hatched background. The gas isotherms of 20 and 40\,K  are shown in blue and red in the upper panels, respectively, highlighting the CO snowline location and shape. 
}
\label{chemical_structure}
\end{figure*}

\begin{figure*}
\centering
     Model \modr{}\hspace{0.25\linewidth} Model \moda{}\hspace{0.25\linewidth} Model \modb{}
\includegraphics[width=\thirdscale\linewidth,clip,page=4]{figures/ref.pdf}
\includegraphics[width=\thirdscale\linewidth,clip,page=4]{figures/gap.pdf}
\includegraphics[width=\thirdscale\linewidth,clip,page=4]{figures/dustgap.pdf}

\includegraphics[width=\thirdscale\linewidth,clip,page=1]{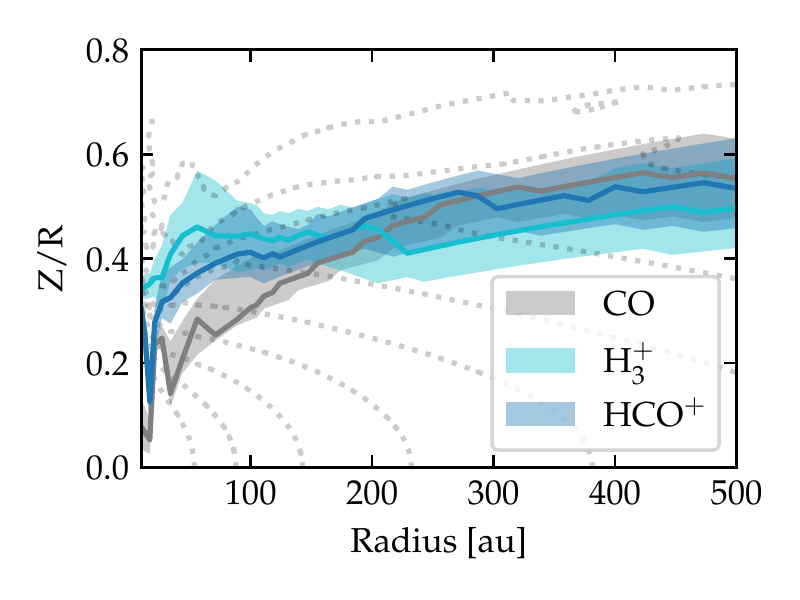}
\includegraphics[width=\thirdscale\linewidth,clip,page=1]{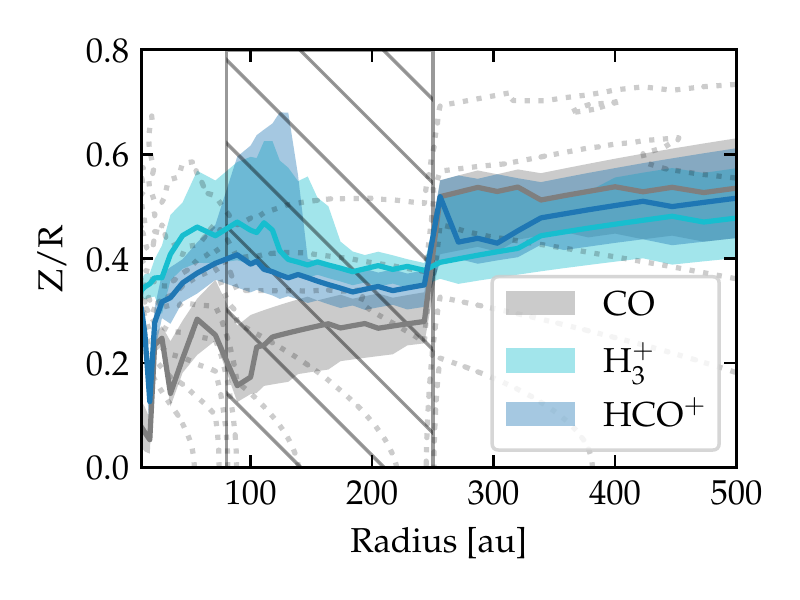}
\includegraphics[width=\thirdscale\linewidth,clip,page=1]{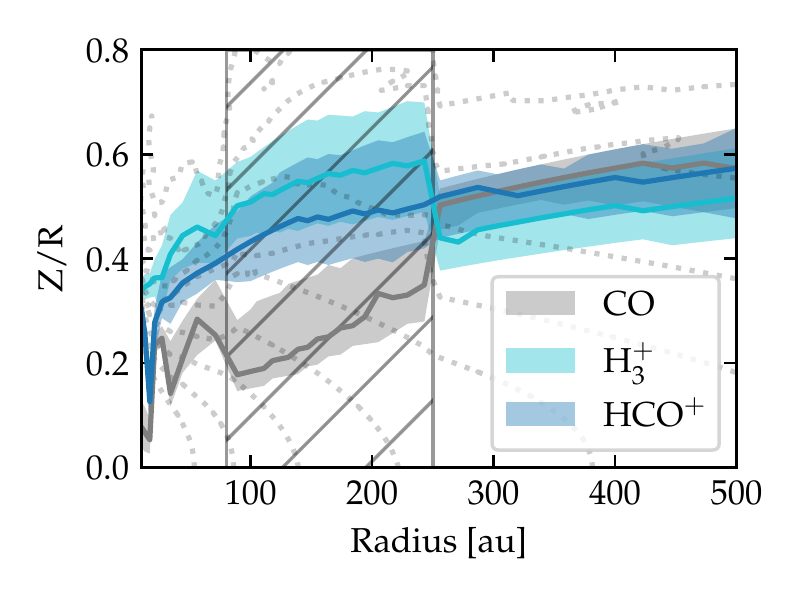}

\includegraphics[width=\thirdscale\linewidth,clip,page=7]{figures/ref.pdf}
\includegraphics[width=\thirdscale\linewidth,clip,page=7]{figures/gap.pdf}
\includegraphics[width=\thirdscale\linewidth,clip,page=7]{figures/dustgap.pdf}

\caption{Vertical mass distribution of \Htp, deuterium in \Htp{} isotopologues, CO, \HCOp{}, and \DCOp{} as a function of the radius. The color-filled stripes show the vertical location of the 25, 50, and 75 mass percentiles (bottom border, median line, and top border, respectively). Half of the total molecular gas mass is located within the corresponding stripe. The dotted contour lines show the gas isodensities from \num{1e-18} to \num{1e-14}~g~cm$^{-2}$ with a logarithmic step of $10^{0.5}$, which is the same as in the top panel of Fig.~\ref{thermal_structure}.
The left panel is the reference model \modr{}, the middle panel is the gas-poor gap model \moda{}, and the right panel is the gas-rich gap model \modb{}. The location of the gap is shown as a hatched background. When molecules are co-spatial, their mass distributions overlap (e.g., as in the case of deuterated isotopologues of \Htp{} in the model \moda{}).
}
\label{deuterium_loc}
\end{figure*}

\begin{figure*}
\centering
\includegraphics[width=0.45\linewidth,clip,page=24]{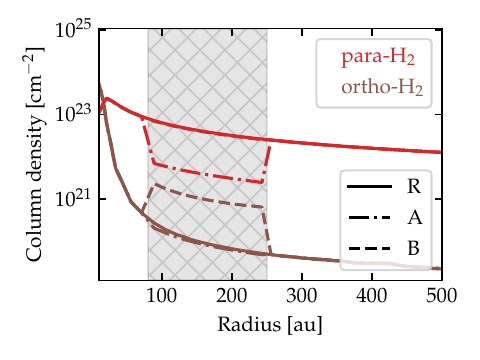}
\includegraphics[width=0.45\linewidth,clip,page=25]{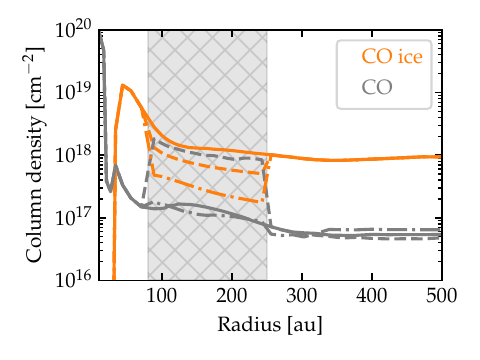}
\includegraphics[width=0.45\linewidth,clip,page=26]{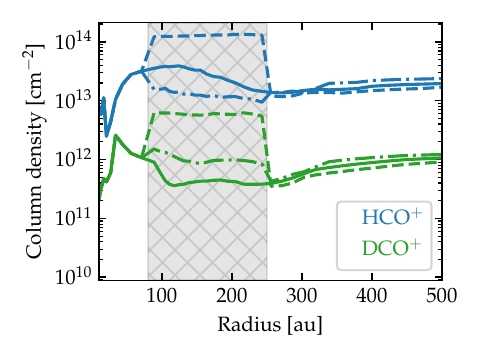}
\includegraphics[width=0.45\linewidth,clip,page=33]{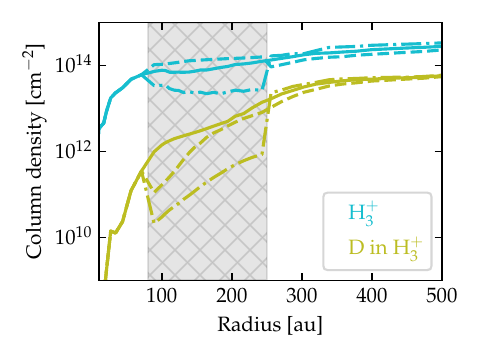}
\caption{Radial profiles of the vertical column densities of the selected molecules.  D in \Htp{} is the total number of D atoms in \Htp{} isotopologues. The solid lines correspond to the reference model \modr{}, the dash-dotted lines to the gas-poor model \moda{}, and the dashed lines to the gas-rich model \modb{}. The location of the gap is shown by the gray rectangle.
}
\label{chem_coldens}
\end{figure*}

\begin{table}
\caption{Initial Chemical Abundances. \label{tab:initial_abundances}}
\begin{tabular*}{\linewidth}{@{\extracolsep{\fill}}lc|lc}
\hline\hline
            \noalign{\smallskip}
        Species & Abundance &Species & Abundance \\
    & $n$(X) / $n$(H) & & $n$(X) / $n$(H) \\ 
            \noalign{\smallskip}
\hline
            \noalign{\smallskip}
ortho-H$_2$ & $3.75\ \times 10^{-1}$ & S        & $9.14\  \times 10^{-8}$         \\
para-H$_2$  & $1.25\ \times 10^ {-1}$ & Si & $9.74\  \times 10^{-9}$    \\
He                      & $9.75\ \times 10^ {-2}$ & Fe & $2.74\  \times 10^{-9}$         \\  
O                       & $1.80\ \times 10^ {-4}$ & Na & $2.25\  \times 10^{-9}$         \\
C                       & $7.86\ \times 10^ {-5}$ & Mg & $1.09\  \times 10^{-8}$         \\  
N                       & $2.47\ \times 10^ {-5}$ & Cl & $1.00\  \times 10^{-9}$         \\
HD                      & $1.55\ \times 10^ {-5}$ & P  & $2.16\  \times 10^{-10}$         \\
            \noalign{\smallskip}
\hline
\end{tabular*}
\end{table}

\subsection{Line radiative transfer}
\label{txt:lime}
   \begin{table}

      \caption[]{LIME input parameters.}
         \label{tab:limesinp}
         \centering 
        \begin{tabular*}{\linewidth}{@{\extracolsep{\fill}}lll}

            \hline \hline
            \noalign{\smallskip}
            Parameter  & Internal name    &  Value \\
            \noalign{\smallskip}
            \hline
            \noalign{\smallskip}
            Minimal spatial scale & minScale & 1\,au \\
            Number of grid points & pIntensity & $10^5$  \\
            Number of sink points & sinkPoints & 3000 \\
            \noalign{\smallskip}
            \hline
            \noalign{\smallskip}
            Velocity resolution & velres & 20 m s$^{-1}$ \\
            Distance to the source & distance & 100 pc\\
            \noalign{\smallskip}
            \hline 
        \end{tabular*}

   \end{table}

To test whether the radial variations in the chemical structures could become visible in the emission images, we have used the computed physical and chemical structures of the disk and performed LTE line radiative transfer with LIME v1.9.4 (L{I}ne Modeling Engine) by \citet{Brinch_Hogerheijde_2010_lime}. We used the corresponding spectroscopic and collisional rate data from the Leiden Atomic and Molecular Database \citep[LAMDA:][]{lamda}. Other key input parameters for LIME are listed in Table \ref{tab:limesinp}.

We tested the LTE line radiative transfer calculations, both with and without the dust opacities taken into account, and we did not notice any significant differences in the continuum-subtracted line emission for the low-J HCO$^+$ and CO isotopologue lines in the studied low-density regions of the disk. As the dust continuum is usually subtracted from the science-ready observational data and since the LIME computations, which include dust opacities, take much longer, we only show the results obtained for the pure gas emission case. For each synthetic image, we ran 50 instances of LIME and then averaged them to reduce the noise. As global emission distributions look similar for the J=5--4, 4--3, and 3--2 transitions, we focus on the J=3--2 results. 

\section{Results}
\label{txt:results}
\begin{figure*}[h!t]
\centering

\includegraphics[page=1,scale=\limescale]{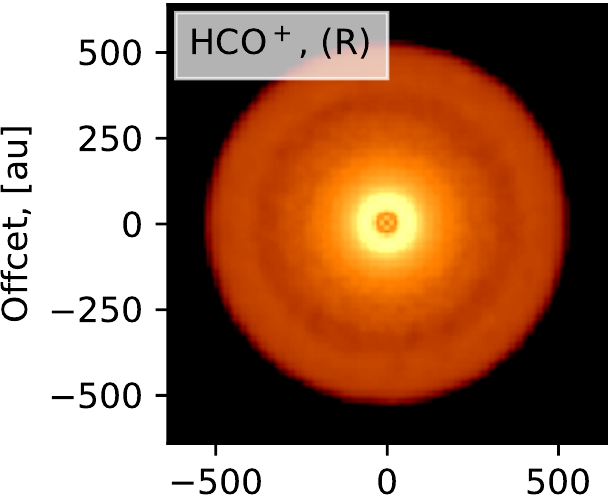}
\includegraphics[page=2,scale=\limescale]{figures/short_cropped.pdf}
\includegraphics[page=3,scale=\limescale]{figures/short_cropped.pdf}
\includegraphics[page=4,scale=\limescale]{figures/short_cropped.pdf}

\includegraphics[page=5,scale=\limescale]{figures/short_cropped.pdf}
\includegraphics[page=6,scale=\limescale]{figures/short_cropped.pdf}
\includegraphics[page=7,scale=\limescale]{figures/short_cropped.pdf}
\includegraphics[page=8,scale=\limescale]{figures/short_cropped.pdf}

\includegraphics[page=9,scale=\limescale]{figures/short_cropped.pdf}
\includegraphics[page=10,scale=\limescale]{figures/short_cropped.pdf}
\includegraphics[page=11,scale=\limescale]{figures/short_cropped.pdf}
\includegraphics[page=12,scale=\limescale]{figures/short_cropped.pdf}

\vspace{0.4cm}

\includegraphics[page=13,scale=\limescale]{figures/short_cropped.pdf}
\includegraphics[page=14,scale=\limescale]{figures/short_cropped.pdf}
\includegraphics[page=15,scale=\limescale]{figures/short_cropped.pdf}
\includegraphics[page=16,scale=\limescale]{figures/short_cropped.pdf}

\includegraphics[page=17,scale=\limescale]{figures/short_cropped.pdf}
\includegraphics[page=18,scale=\limescale]{figures/short_cropped.pdf}
\includegraphics[page=19,scale=\limescale]{figures/short_cropped.pdf}
\includegraphics[page=20,scale=\limescale]{figures/short_cropped.pdf}

\caption{Integrated intensity 0th-moment maps. From left to right: \HCOp{}, \DCOp{}, \HnCOp{}, and C\element[][18][][]{O}, all J=3--2. From the top to the bottom row: the reference model \modr{}, the gas-poor gap model \moda{}, the gas-rich model \modb{}, the ratio \moda{}/\modr{}, and the ratio \modb{}/\modr{}. The gap boundaries are marked by the two dotted lines. The optically thick \HCOp{} emission is not significantly affected by the presence of the gas-poor gap. In contrast,  the C\element[][18][][]{O} and particularly \HnCOp{} intensities decrease inside the gas-poor gap, while the \DCOp{} intensity increases inside this gap. 
}
\label{isotopologues_lime}
\end{figure*}

\subsection{Reference model \modr{}}
\label{txt:res-r}
The reference model does not have any density gaps. Its gas surface density is set by the the power law (Eq.~\ref{eq:coldens}) with $\Sigma_{1 \rm{au}} = 100\ \rm {g\ cm}^{-1}$. This type of monotonous surface density profile still leads to rather nonmonotonous abundance structures (see Fig.~\ref{chemical_structure}). This nonhomogeneous abundance distribution is typical of disk models with gas-grain chemistry, which are sensitive to the local variations in temperature, density, ionization, and high-energy radiation intensities \citep[e.g.,][]{Semenov_Wiebe11}. 

In our reference disk model, the self-shielded CO extends {far} into the disk atmosphere,  until the height-to-radius ratio of $\sim 0.5$ and above (Figs.~\ref{chemical_structure} \& \ref{deuterium_loc}).
The CO snowline is located at about 30~au in the midplane. Inside the radius of 20 au, the gas-phase CO exists all the way through the disk down to the midplane. The CO molecules are efficiently converted to CO$_2$ in a region between $\sim 20-30$~au, causing the gas-phase CO abundances to decrease at height-to-radius ratio below $\sim 0.2$. Outside of 30~au, the gas-phase CO resides in the molecular layer above the height and radius scales of $\sim 0.1-0.3$. The CO ice is mainly concentrated in the midplane at $r \gtrsim 30$~au, and the height-to-radius ratio  below $\sim  0.3$.

The distribution of the H$_3^+$ isotopologues sensitively depends on the local ionization rate and the local gas density, since its primary production mechanism involves the ionization of H$_2$ followed by the ion-molecule reactions of H$_2^+$ with H$_2$. Consequently, in the dense midplane, where ionization rates are low,
the abundances of H$_3^+$ isotopologues decrease, but they never disappear entirely as they do for the gas-phase CO. A  key difference between the main H$_3^+$ isotopologue and its  minor D-isotopologues is that the molecular layer of the deuterated H$_3^+$ ions is shifted closer to the cold disk midplane compared to that of H$_3^+$ (Fig.~\ref{deuterium_loc}, left panels). This is because unlocking D from HD via deuterium exchange reactions most rapidly proceeds at temperatures below $\sim 20-100$~K and is particularly efficient when CO is frozen out and when H$_2$ mainly exists in the para form \citep[e.g.,][]{2006A&A...449..621F,Albertsson_ea13,2013A&A...554A..92S,Teague_ea_2015,Huang2017}. This leads to a situation when abundances of, for example, H$_2$D$^+$ can become comparable to the H$_3^+$ abundances in some disk locations, while their vertically-integrated column density would still differ by a factor of $\sim 10-100$ (Fig.~\ref{chemical_structure} and \ref{chem_coldens}, bottom left panels). {Panels with the label "D in \Htp{}" show the total number of D atoms in \Htp{} isotopologues (H$_2$D$^+$, HD$_2^+$, and D$_3^+$).}

The global distribution of the \HCOp{} isotopologues is
determined by the distribution of their parental species, the gaseous CO and the H$_3^+$ isotopologues.   
In general, the \HCOp{} and \DCOp{} abundances do follow the H$_3^+$ and CO gas-phase distributions and they are absent in the CO depletion region. However, there are still major differences between these ions.  First,  the \DCOp{} molecular layer is, by necessity, co-spatial with the upper part of the H$_3^+$ deuterated isotopologue layer, where CO is still not completely frozen out (Fig.~\ref{deuterium_loc}, top left panel). Second, because of that, the \DCOp{} layer is located beneath the \HCOp{} layer and thus probes slightly different physical conditions in the disk (Fig.~\ref{deuterium_loc}, bottom left panel).

As a result of localized variations in the chemical structure, the vertical column densities of the \HCOp{} isotopologues and other key species in the gap-free reference model do show the presence of weak chemical gaps at outer radii or $r > 100$~au (Fig.~\ref{chem_coldens}). Furthermore, these chemical gaps affect the line excitation and appear as emission gaps on the ideal synthetic spectra for the reference disk model with the monotonous global physical 
structure (see Fig.~\ref{isotopologues_lime}). The appearance of these chemical gaps is different for various \HCOp{} isotopologues and C\element[][18][][]{O}, and it depends on their line optical depths and/or the location of the emitting layer. Thus, when taking molecular emission gaps at face value, one could misinterpret the data as being indicative of real physical gaps in the disk gas, while it may not be the case.

\subsection{Gas-poor model \moda{}}
\label{txt:res-c}

The lower amount of dust and gas leads to higher temperatures, larger pressure scale heights, and hence lower densities in the gas-poor gap model compared to the reference model (Fig.~\ref{thermal_structure}, middle panels). The reduced opaqueness of the disk matter in the gas-poor gap results in a vertical shift of the molecular layers down toward the midplane (Figs.~\ref{chemical_structure} and \ref{deuterium_loc}, middle panels). While the vertical column density of H$_2$ and H$_3^+$ and the total (ice and gas) CO concentration decrease by a factor of $\sim 10$ inside the gap, the vertical column density of gaseous CO almost remains intact. The balance between the CO photodissociation and freeze-out is similar in the shifted CO layer inside the gap, just as in the gap-free reference model (Fig.~\ref{chem_coldens}, top right panel). Naturally, the total amount of CO ice, which dominates CO concentration outside the CO snowline, is lower in the gas-poor gap model \moda{} compared to the reference model \modr. 

This vertical shift of the CO molecular layer in the gas-poor gap brings it closer to the molecular layer of the \Htp{} D-isotopologues,
but moves it away from the main \Htp{} molecular layer (Fig.~\ref{deuterium_loc}, middle panels). 
It leads to an increase in the rate of the \DCOp{}-forming reaction, which is proportional to a product of deuterated isotopologues of \Htp{} and CO volume densities. However, it also leads to an opposite effect for the \HCOp{} abundances. 
Consequently, 
the column density of \HCOp{} decreases by a factor of $\sim 3$, whereas the column density of \DCOp{} increases by a similar factor (Fig.~\ref{chem_coldens}, bottom left panel). 

These chemical effects remain visible in the emission maps. Therefore, the gas-poor gap model leads to weaker \HnCOp{} emission inside the gap, while the chemically-unaffected, optically thin  C\element[][18][][]{O} and optically-thick \HCOp{} emission remain relatively unaffected. In contrast, the \DCOp{} emission increases inside the gas-poor gap (Fig.~\ref{isotopologues_lime}). This effect is very similar to the observed behavior of \DCOp{},  \HnCOp{}, and C\element[][18][][]{O} in AS~209 \citep{Huang2017,Favre_ea_2019_AS209}. 

\subsection{Gas-rich, dust-poor model \modb{}}
\label{txt:res-d}
Similarly to the gas-poor model \moda{}, the disk model where the gap is only devoid from dust, model \modb{} shows higher temperatures and slightly lower volume densities compared to the reference model \modr{} (Fig.~\ref{thermal_structure}, right panels). The {smaller dust column density} in the gas-rich gap leads to lower extinction of the ionizing and dissociating radiation and makes molecular freeze-out less efficient. This also shifts the molecular layers closer to the midplane, as in the previous gap model (Figs.~\ref{chemical_structure} and \ref{deuterium_loc}, right panels). 

The dust-depleted gap with the same amount of gas as in the reference model leads to about a 10 times higher column density of gaseous CO and a $\sim 2-3$ times lower column density of the CO ice (Fig.~\ref{chem_coldens}). The higher concentration of gaseous CO in the cold midplane region makes reactions with the \Htp D-isotopologues faster, while a higher concentration of CO in the molecular layer also boosts the production of \HCOp{}. Thus, the model with the dust-poor gap leads to a  similar increase in the column densities of \HCOp{}, \DCOp{}, and CO. This strong increase in the molecular concentrations is imprinted into the corresponding increase in the \HCOp{}, \DCOp{}, and C\element[][18][][]{O} emission in the dust-poor gap (Fig.~\ref{isotopologues_lime}). Thus, using the single ALMA Band~6 or NOEMA Band~3 spectral setups with \DCOp{} and the CO isotopologue or the two setups with the \HCOp{}, \HnCOp{}, and \DCOp{} ions, one could distinguish between gas-poor and dust-poor disk gaps and estimate the overall depletion of gas and/or dust there. 



\section{Discussion}
\label{txt:discussions}
\subsection{Model uncertainties and future development}
\label{txt:uncertainties}

The version of ANDES used in this study simulates each disk radius independently, thus limiting us in the modeling approaches for ionizing radiation transport that is consistent with the thermal structure. The X-rays and stellar energetic particles from the star cause significant heating in the gas, affecting its vertical structure and ionization fraction. The 2D nature of a gap must have an impact on the underlying chemical structure and thus it should also affect the gas thermal balance. 
Consistent 2D modeling with accurate modeling of the gas thermal balance will be implemented in the future 2D version of {ANDES$_{\rm 2D}$}, first presented in \citet{2017ApJ...849..130M, 2018ApJ...866...46M}, which is being actively developed. 

From the chemical point of view, the calculated abundances of simple species in our model suffer from intrinsic uncertainties due to the reaction rate uncertainties, which are on the order of a factor of 3 for \HCOp{} isotopologues and less than a factor of 2 for the gas-phase CO \citep{Vasyunin_ea08,Albertsson_ea13,Iqbal_ea18}. 
In addition, our chemical network does not include reactions involving isotopologues except for deuterium, and hence we had to rescale the abundances of HCO$^+$ and CO to
get the \HnCOp{} and C\element[][18][][]{O} abundances. This simplification may introduce additional chemical uncertainty by a factor of 2-3 for both these minor isotopologues. The main reason for that is the computational demands required for our model with deuterium fractionation. A further increase in the number of species from $\sim 1250$ in the current version to $>2000-5000$ species in the network that also includes key \element[][13]C- and \element[][18]O-isotopologues would {increase the computation time of the chemical evolution by at least a factor of 10}, and we decided to leave the more detailed analysis for future studies.


\subsection{Comparison with previous studies}
\label{txt:comparison}
\citet{2018A&A...614A.106C} have studied the gap at 40--60\,au between the two dusty rings in the protoplanetary disk around HD~169142. In contrast to our case, \DCOp{} J=3--2 emission intensity decreases in this gap. The thermo-chemical structure of the HD~169142 disk was simulated with the DALI code \citep{Bruderer_ea_2014_DALI2}. To reproduce the observed behavior, \citet{Bruderer_ea_2014_DALI2} assumed the low gas densities in both the gap and the inner dust ring, similar to our gas-poor model~\moda{}. While the authors do not show the \HCOp{} radial profiles, the behavior of the \DCOp{} emission alone is neither consistent with the \DCOp{} observations of AS~209 nor with our modeling. The first reason for this inconsistency could be a stronger gas density depletion, {estimated as} a factor of 40 in their model, in comparison to our factor of 10 depletion. In their case, the \DCOp{}~/~\element[][][][2]{H} ratio inside the gap is lower by a factor of $\sim 20$ in comparison with the outer dust ring.
This is somewhat similar to the increase we have in the gas-poor model \moda{}. The second major difference is that this is a Herbig~Ae disk (the stellar mass is 1.65 $M_\odot$).
Compared to our cold T~Tauri disk with the CO snowline located at $\sim 30$~au, the CO snowline in the HD~169142 disk is likely located beyond 50~au. Thus, the gap at 40--60\,au in 
HD~169142 could be inside the CO snowline and not outside, as in our model. 

\begin{figure}[h]
\centering

\includegraphics[width=0.45\textwidth,clip,page=39]{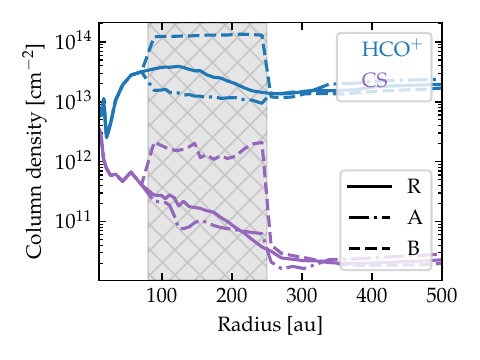}
\includegraphics[width=0.45\textwidth,clip,page=40]{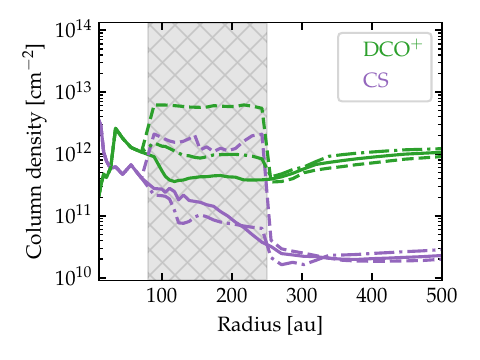}
\caption{Column densities of CS (purple) in comparison with the \HCOp{} (blue) and \DCOp{} (green) column densities. The solid line denotes the reference model \modr{}, the dash-dotted line is the gas-poor model \moda{}, and the dashed line is the gas-rich model \modb{}. The location of the gap is shown by the gray box.
}
\label{cs_coldens}
\end{figure}

\citet{Teague_ea_2017} have studied the dip in the CS emission at the radius of $r\sim 80-90$~au in TW~Hya, which coincides with one of the (sub)mm dust gaps in this disk. They computed a set of models with different maximal depths of the Gaussian gap ($\Sigma_x / \Sigma_A = 0.3 (B),\ 0.55 (C),\ 1 (D)$). They considered several cases of the gap, both with and without gas inside. The model that qualitatively reproduces the observed CS radial profile, especially its second derivative, was model \textit{C} with an intermediate gap depth. The behavior of the CS emission in TW~Hya is similar to that of \HCOp{} in our model. In Fig.~\ref{cs_coldens} we show the CS column density derived from our model in comparison with \HCOp{} and \DCOp{}. CS is more sensitive to the lowering of the dust-to-gas ratio because it unlocks more carbon from CO by photodissociation and due to less efficient freeze out. In the gas-poor gap model \moda{}, there is no significant difference in the CS abundance with respect to the reference model \modr{}. On the other hand, in the case of the gas-rich gap model \modb{}, the CS abundance increases stronger than the abundance of \HCOp{} with respect to the reference model \modr{}. It is also consistent with the results of \citet{Teague_ea_2017} for their gas-rich model \textit{Cd}. Thus, CS, together with \DCOp{}, can also serve as a probe for gas depletion in disk gaps if \HnCOp{} data are not available.

\begin{figure}[h]
\centering

\includegraphics[width=0.45\textwidth,clip]{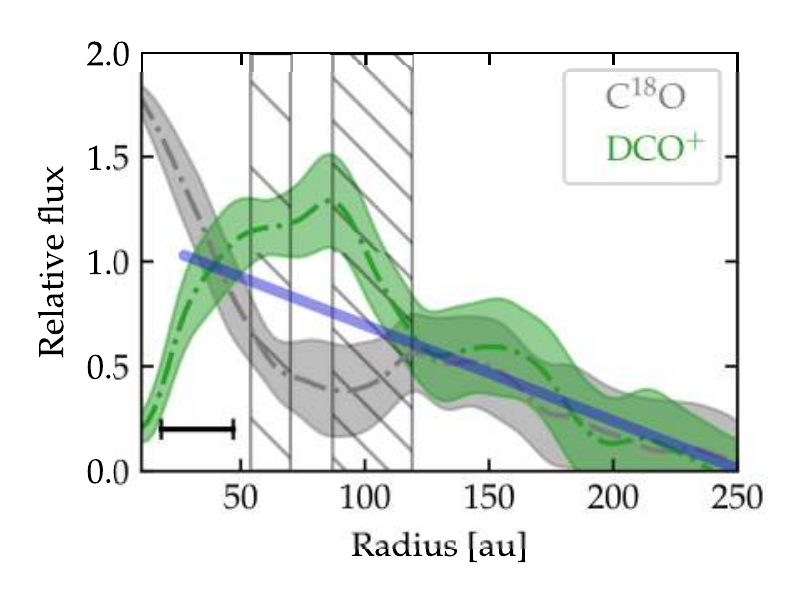}
\caption{{Azimuthally averaged  profiles of \DCOp{} (green) and C$^{18}$O (gray) emission line fluxes from \citet{Favre_ea_2019_AS209}. The profiles were normalized to have the same slope outside of the gap (120 - 250 au); the slope is shown in blue. There is an increase in \DCOp{} emission and a decrease in C$^{18}$O emission intensity at the gap location in comparison with it.  In the bottom left corner of the image, the beam size of 0.23'' is shown. The locations of the gaps are shown by the hatched boxes.
}}
\label{favre_profile}
\end{figure}

Finally, \citet{Favre_ea_2019_AS209} performed the thermo-chemical modeling to fit the observed CO isotopologues emission in AS~209. For the outer gap at $\sim 100$~au, they modeled the depletion of both gas and dust. Unlike {this work}, they assumed that the dust surface density is depleted by a factor of 100, while the gas surface density is depleted by only a factor of $2.5-10$. Our results are consistent with their modeling results, as the stronger gas depletion reproduces the CO observations in the gap better. While they have shown the ALMA data for the \DCOp{} J=3--2 emission, the modeling involving deuterated chemistry has been left for the forthcoming paper. {The azimuthally averaged \DCOp{} and C$^{18}$O emission line profiles and the location of the assumed gaps are shown in Fig.~\ref{favre_profile}. The profiles were normalized to have the same slope in the outer disk. The increase in the \DCOp{} flux and the decrease in C$^{18}$O in comparison to their outer disk slope at the location of the gaps are clearly visible, as in our gas-poor model \moda{} (Fig.~\ref{isotopologues_lime}, second and fourth rows).}

{However, our modeling does not fit the AS~209 observations quantitatively.} The stellar mass of $0.5\ M_\odot{}$ and thus the relatively low luminosity that we adopted in our models are typical for {low-mass} T~Tauri stars. However, the type K4-5 AS~209 star is more luminous and massive and its disk is likely warmer than our {disk} models \citep{2018ApJ...868..113T}. Compared to our disk models, the CO snowline in the AS~209 disk is located at $\sim 30-40$~au (or about 10~au farther out from the central star). Still, it does not change the main conclusion of our study because the outer gap in AS~209 is  located outside of the CO snowline, as in our models, leading to similar  chemical effects for the \HCOp{} isotopologues.
\section{Conclusion}
\label{txt:conclusion}
We present detailed physical-chemical and line radiative transfer simulations of protoplanetary disks, both with and without a wide gap. For that, we used the ANDES thermo-chemical model to calculate 1+1D thermal and density disk structures and included various ionization sources (FUV, X-rays, SEP, CRs, and SLRs). Next, we used the gas-grain chemical code ALCHEMIC with a deuterated network to calculate molecular abundances in the disk. Finally, we utilized LIME for the line radiative transfer to produce the ideal synthetic images in  molecular lines of the \HCOp{} isotopologues and C\element[][18][][]{O}. Three disk models were considered: the reference gap-free model, the model with the gas-poor gap where both gas and dust surface densities are depleted by a factor of 10, and the gas-rich gap model where solely the dust surface density is depleted by a factor of 10.  

Our simulations reveal that in using the \HCOp{} and CO isotopologues, one could observationally distinguish between various types of the disk gaps (gas-poor versus gas-rich). 
The most convenient proxy of the gas or dust depletion in the disk gaps is either the \DCOp{}/\HnCOp{} and \DCOp{}/\HCOp{} line ratios (using two ALMA Band~6 or NOEMA Band~3 setups) or the \DCOp{}/C\element[][18][][]{O} line ratio (using single ALMA Band~6 or NOEMA Band~3 spectral setups). Together with high-resolution dust continuum 
data, these ratios allow one to estimate the degree of the gas or dust depletion in the disk gaps. Namely, if most of the emission lines appear brighter inside the (sub)mm dust gap, especially \HnCOp{}, it means that the gas in the gap is not strongly depleted. In contrast, if C\element[][18][][]{O} or \HnCOp emission decreases while the \DCOp{}
emission increases inside the gap, it means that the disk gap is substantially depleted in both gas and dust.
Finally, the results of our study using the gas-poor disk model are in accordance with the ALMA observations of AS~209, where C\element[][18][][]{O} and \DCOp{} emission inside the outer $\sim 100$~au gap shows the opposite behavior. Thus, using intensity ratios of \DCOp{} and C\element[][18][][]{O} or the \HCOp{} isotopologues, one might get invaluable information about the disk gap physics and the gap clearing mechanisms.

\begin{acknowledgements} 
The authors thank Cecile Favre for sharing the azimuthally averaged emission line profiles for AS~209 ALMA observations. The authors acknowledge the Python open-source community. This work was done using the open-source packages NumPy, SciPy, AstroPy, and Matplotlib. D.S. acknowledges support by the Deutsche Forschungsgemeinschaft through SPP 1833:
``Building a Habitable Earth'' (SE 1962/6-1).
{
V.A. acknowledges the support of the Large Scientific Project of the Russian Ministry of Science and Higher Education "Theoretical and experimental studies of the formation and evolution of extrasolar planetary systems and characteristics of exoplanets" (project No. 13.1902.21.0039).}
T.H. acknowledges support from the European Research Council under the
Horizon 2020 Framework Program via the ERC Advanced Grant Origins 83 24 28.
This research has made use of NASA’s Astrophysics Data System Bibliographic Services.
\end{acknowledgements}

\bibliography{bibliography}{}

\begin{thebibliography}{90}
\expandafter\ifx\csname natexlab\endcsname\relax\def\natexlab#1{#1}\fi

\bibitem[{{Akimkin} {et~al.}(2013){Akimkin}, {Zhukovska}, {Wiebe}, {Semenov},
  {Pavlyuchenkov}, {Vasyunin}, {Birnstiel}, \& {Henning}}]{Akimkin_ea_2013}
{Akimkin}, V., {Zhukovska}, S., {Wiebe}, D., {et~al.} 2013, \apj, 766, 8

\bibitem[{{Albertsson} {et~al.}(2014){Albertsson}, {Semenov}, \&
  {Henning}}]{Albertsson_ea14a}
{Albertsson}, T., {Semenov}, D., \& {Henning}, T. 2014, ApJ, 784, 39

\bibitem[{{Albertsson} {et~al.}(2013){Albertsson}, {Semenov}, {Vasyunin},
  {Henning}, \& {Herbst}}]{Albertsson_ea13}
{Albertsson}, T., {Semenov}, D.~A., {Vasyunin}, A.~I., {Henning}, T., \&
  {Herbst}, E. 2013, \apjs, 207, 27

\bibitem[{{Andrews} {et~al.}(2018){Andrews}, {Huang}, {P{\'e}rez}, {Isella},
  {Dullemond}, {Kurtovic}, {Guzm{\'a}n}, {Carpenter}, {Wilner}, {Zhang}, {Zhu},
  {Birnstiel}, {Bai}, {Benisty}, {Hughes}, {{\"O}berg}, \& {Ricci}}]{DSHARP1}
{Andrews}, S.~M., {Huang}, J., {P{\'e}rez}, L.~M., {et~al.} 2018, \apjl, 869,
  L41

\bibitem[{{Bai} \& {Goodman}(2009)}]{Bai_Goodman_2009}
{Bai}, X.-N. \& {Goodman}, J. 2009, \apj, 701, 737

\bibitem[{{Bergin} {et~al.}(2016){Bergin}, {Du}, {Cleeves}, {Blake}, {Schwarz},
  {Visser}, \& {Zhang}}]{Bergin_ea16}
{Bergin}, E.~A., {Du}, F., {Cleeves}, L.~I., {et~al.} 2016, \apj, 831, 101

\bibitem[{{Biham} {et~al.}(2001){Biham}, {Furman}, {Pirronello}, \&
  {Vidali}}]{Bihamea01}
{Biham}, O., {Furman}, I., {Pirronello}, V., \& {Vidali}, G. 2001, \apj, 553,
  595

\bibitem[{{Brinch} \& {Hogerheijde}(2010)}]{Brinch_Hogerheijde_2010_lime}
{Brinch}, C. \& {Hogerheijde}, M.~R. 2010, \aap, 523, A25

\bibitem[{{Bruderer} {et~al.}(2009){Bruderer}, {Doty}, \&
  {Benz}}]{Bruderer_ea_2009a}
{Bruderer}, S., {Doty}, S.~D., \& {Benz}, A.~O. 2009, \apjs, 183, 179

\bibitem[{{Bruderer} {et~al.}(2014){Bruderer}, {van der Marel}, {van Dishoeck},
  \& {van Kempen}}]{Bruderer_ea_2014_DALI2}
{Bruderer}, S., {van der Marel}, N., {van Dishoeck}, E.~F., \& {van Kempen},
  T.~A. 2014, \aap, 562, A26

\bibitem[{{Carney} {et~al.}(2018){Carney}, {Fedele}, {Hogerheijde}, {Favre},
  {Walsh}, {Bruderer}, {Miotello}, {Murillo}, {Klaassen}, {Henning}, \& {van
  Dishoeck}}]{2018A&A...614A.106C}
{Carney}, M.~T., {Fedele}, D., {Hogerheijde}, M.~R., {et~al.} 2018, \aap, 614,
  A106

\bibitem[{{Cazzoletti} {et~al.}(2018){Cazzoletti}, {van Dishoeck}, {Visser},
  {Facchini}, \& {Bruderer}}]{2018A&A...609A..93C}
{Cazzoletti}, P., {van Dishoeck}, E.~F., {Visser}, R., {Facchini}, S., \&
  {Bruderer}, S. 2018, \aap, 609, A93

\bibitem[{{Ceccarelli} {et~al.}(2014){Ceccarelli}, {Caselli},
  {Bockel{\'e}e-Morvan}, {Mousis}, {Pizzarello}, {Robert}, \&
  {Semenov}}]{Ceccarelli_ea14}
{Ceccarelli}, C., {Caselli}, P., {Bockel{\'e}e-Morvan}, D., {et~al.} 2014, in
  Protostars and Planets VI, ed. H.~{Beuther}, R.~S. {Klessen}, C.~P.
  {Dullemond}, \& T.~{Henning}, 859

\bibitem[{{Cleeves} {et~al.}(2013){Cleeves}, {Adams}, \&
  {Bergin}}]{Cleeves_ea_2013}
{Cleeves}, L.~I., {Adams}, F.~C., \& {Bergin}, E.~A. 2013, \apj, 772, 5

\bibitem[{{Cleeves} {et~al.}(2015){Cleeves}, {Bergin}, \&
  {Harries}}]{Cleeves_ea15}
{Cleeves}, L.~I., {Bergin}, E.~A., \& {Harries}, T.~J. 2015, \apj, 807, 2

\bibitem[{{Cleeves} {et~al.}(2017){Cleeves}, {Bergin}, {{\"O}berg}, {Andrews},
  {Wilner}, \& {Loomis}}]{Cleeves_ea17}
{Cleeves}, L.~I., {Bergin}, E.~A., {{\"O}berg}, K.~I., {et~al.} 2017, \apjl,
  843, L3

\bibitem[{{Dolginov} \& {Stepinski}(1994)}]{DS_94}
{Dolginov}, A.~Z. \& {Stepinski}, T.~F. 1994, \apj, 427, 377

\bibitem[{{Draine} \& {Lee}(1984)}]{1984ApJ...285...89D}
{Draine}, B.~T. \& {Lee}, H.~M. 1984, \apj, 285, 89

\bibitem[{{Du} \& {Bergin}(2014)}]{2014ApJ...792....2D}
{Du}, F. \& {Bergin}, E.~A. 2014, \apj, 792, 2

\bibitem[{{Dutrey} {et~al.}(2017){Dutrey}, {Guilloteau}, {Pi{\'e}tu},
  {Chapillon}, {Wakelam}, {Di Folco}, {Stoecklin}, {Denis-Alpizar}, {Gorti},
  {Teague}, {Henning}, {Semenov}, \& {Grosso}}]{Dutrey_ea17}
{Dutrey}, A., {Guilloteau}, S., {Pi{\'e}tu}, V., {et~al.} 2017, \aap, 607, A130

\bibitem[{{Favre} {et~al.}(2019){Favre}, {Fedele}, {Maud}, {Booth}, {Tazzari},
  {Miotello}, {Testi}, {Semenov}, \& {Bruderer}}]{Favre_ea_2019_AS209}
{Favre}, C., {Fedele}, D., {Maud}, L., {et~al.} 2019, \apj, 871, 107

\bibitem[{{Fedele} {et~al.}(2017){Fedele}, {Carney}, {Hogerheijde}, {Walsh},
  {Miotello}, {Klaassen}, {Bruderer}, {Henning}, \& {van
  Dishoeck}}]{2017A&A...600A..72F}
{Fedele}, D., {Carney}, M., {Hogerheijde}, M.~R., {et~al.} 2017, \aap, 600, A72

\bibitem[{{Fedele} {et~al.}(2018){Fedele}, {Tazzari}, {Booth}, {Testi},
  {Clarke}, {Pascucci}, {Kospal}, {Semenov}, {Bruderer}, {Henning}, \&
  {Teague}}]{2018A&A...610A..24F}
{Fedele}, D., {Tazzari}, M., {Booth}, R., {et~al.} 2018, \aap, 610, A24

\bibitem[{{Flower} {et~al.}(2004){Flower}, {Pineau des For{\^e}ts}, \&
  {Walmsley}}]{2004A&A...427..887F}
{Flower}, D.~R., {Pineau des For{\^e}ts}, G., \& {Walmsley}, C.~M. 2004, \aap,
  427, 887

\bibitem[{{Flower} {et~al.}(2006){Flower}, {Pineau Des For{\^e}ts}, \&
  {Walmsley}}]{2006A&A...449..621F}
{Flower}, D.~R., {Pineau Des For{\^e}ts}, G., \& {Walmsley}, C.~M. 2006, \aap,
  449, 621

\bibitem[{{Gammie}(1996)}]{Gammie_96}
{Gammie}, C.~F. 1996, \apj, 457, 355

\bibitem[{{Garrod} \& {Herbst}(2006)}]{Garrod_Herbst06}
{Garrod}, R.~T. \& {Herbst}, E. 2006, \aap, 457, 927

\bibitem[{{Garrod} {et~al.}(2007){Garrod}, {Wakelam}, \&
  {Herbst}}]{2007A&A...467.1103G}
{Garrod}, R.~T., {Wakelam}, V., \& {Herbst}, E. 2007, \aap, 467, 1103

\bibitem[{{Garufi} {et~al.}(2020){Garufi}, {Podio}, {Codella}, {Rygl},
  {Bacciotti}, {Facchini}, {Fedele}, {Miotello}, {Teague}, \&
  {Testi}}]{Garufi_ea20a}
{Garufi}, A., {Podio}, L., {Codella}, C., {et~al.} 2020, \aap, 636, A65

\bibitem[{{Gerlich}(1990)}]{1990JChPh..92.2377G}
{Gerlich}, D. 1990, \jcp, 92, 2377

\bibitem[{{Gerlich} {et~al.}(2002){Gerlich}, {Herbst}, \& {Roueff}}]{GHR_02}
{Gerlich}, D., {Herbst}, E., \& {Roueff}, E. 2002, Planet.~Space~Sci., 50, 1275

\bibitem[{{Gorti} \& {Hollenbach}(2004)}]{2004ApJ...613..424G}
{Gorti}, U. \& {Hollenbach}, D. 2004, \apj, 613, 424

\bibitem[{{Guzm{\'a}n} {et~al.}(2018){Guzm{\'a}n}, {Huang}, {Andrews},
  {Isella}, {P{\'e}rez}, {Carpenter}, {Dullemond}, {Ricci}, {Birnstiel},
  {Zhang}, {Zhu}, {Bai}, {Benisty}, {{\"O}berg}, \& {Wilner}}]{Guzman_ea18a}
{Guzm{\'a}n}, V.~V., {Huang}, J., {Andrews}, S.~M., {et~al.} 2018, \apjl, 869,
  L48

\bibitem[{{Harada} {et~al.}(2010){Harada}, {Herbst}, \&
  {Wakelam}}]{Harada_ea10}
{Harada}, N., {Herbst}, E., \& {Wakelam}, V. 2010, Astrophys.~J.,, 721, 1570

\bibitem[{{Harada} {et~al.}(2012){Harada}, {Herbst}, \&
  {Wakelam}}]{Harada_ea12}
{Harada}, N., {Herbst}, E., \& {Wakelam}, V. 2012, \apj, 756, 104

\bibitem[{{Hasegawa} {et~al.}(1992){Hasegawa}, {Herbst}, \& {Leung}}]{HHL92}
{Hasegawa}, T.~I., {Herbst}, E., \& {Leung}, C.~M. 1992, \apjs, 82, 167

\bibitem[{{Honvault} {et~al.}(2011){Honvault}, {Jorfi}, {Gonz{\'a}lez-Lezana},
  {Faure}, \& {Pagani}}]{2011PhRvL.107b3201H}
{Honvault}, P., {Jorfi}, M., {Gonz{\'a}lez-Lezana}, T., {Faure}, A., \&
  {Pagani}, L. 2011, Physical Review Letters, 107, 023201

\bibitem[{{Huang} {et~al.}(2018{\natexlab{a}}){Huang}, {Andrews}, {Cleeves},
  {{\"O}berg}, {Wilner}, {Bai}, {Birnstiel}, {Carpenter}, {Hughes}, {Isella},
  {P{\'e}rez}, {Ricci}, \& {Zhu}}]{Huang_ea_2018_TWHya}
{Huang}, J., {Andrews}, S.~M., {Cleeves}, L.~I., {et~al.} 2018{\natexlab{a}},
  \apj, 852, 122

\bibitem[{{Huang} {et~al.}(2018{\natexlab{b}}){Huang}, {Andrews}, {Dullemond},
  {Isella}, {P{\'e}rez}, {Guzm{\'a}n}, {{\"O}berg}, {Zhu}, {Zhang}, {Bai},
  {Benisty}, {Birnstiel}, {Carpenter}, {Hughes}, {Ricci}, {Weaver}, \&
  {Wilner}}]{DSHARP2}
{Huang}, J., {Andrews}, S.~M., {Dullemond}, C.~P., {et~al.} 2018{\natexlab{b}},
  \apjl, 869, L42

\bibitem[{{Huang} {et~al.}(2020){Huang}, {Andrews}, {Dullemond}, {{\"O}berg},
  {Qi}, {Zhu}, {Birnstiel}, {Carpenter}, {Isella}, {Mac{\'\i}as}, {McClure},
  {P{\'e}rez}, {Teague}, {Wilner}, \& {Zhang}}]{Huang_ea20a}
{Huang}, J., {Andrews}, S.~M., {Dullemond}, C.~P., {et~al.} 2020, \apj, 891, 48

\bibitem[{{Huang} \& {{\"O}berg}(2015)}]{Huang_Oberg15}
{Huang}, J. \& {{\"O}berg}, K.~I. 2015, \apjl, 809, L26

\bibitem[{Huang {et~al.}(2017)Huang, {\"{O}}berg, Qi, Aikawa, Andrews, Furuya,
  Guzm{\'{a}}n, Loomis, {Van Dishoeck}, \& Wilner}]{Huang2017}
Huang, J., {\"{O}}berg, K.~I., Qi, C., {et~al.} 2017, \aj, 835

\bibitem[{{Hugo} {et~al.}(2009){Hugo}, {Asvany}, \&
  {Schlemmer}}]{2009JChPh.130p4302H}
{Hugo}, E., {Asvany}, O., \& {Schlemmer}, S. 2009, \jcp, 130, 164302

\bibitem[{{Igea} \& {Glassgold}(1999)}]{Igea_Glassgold_1999}
{Igea}, J. \& {Glassgold}, A.~E. 1999, \apj, 518, 848

\bibitem[{{Iqbal} {et~al.}(2018){Iqbal}, {Wakelam}, \& {Gratier}}]{Iqbal_ea18}
{Iqbal}, W., {Wakelam}, V., \& {Gratier}, P. 2018, \aap, 620, A109

\bibitem[{{Isella} {et~al.}(2016){Isella}, {Guidi}, {Testi}, {Liu}, {Li}, {Li},
  {Weaver}, {Boehler}, {Carperter}, {De Gregorio-Monsalvo}, {Manara}, {Natta},
  {P{\'e}rez}, {Ricci}, {Sargent}, {Tazzari}, \&
  {Turner}}]{2016PhRvL.117y1101I}
{Isella}, A., {Guidi}, G., {Testi}, L., {et~al.} 2016, \prl, 117, 251101

\bibitem[{{Keppler} {et~al.}(2019){Keppler}, {Teague}, {Bae}, {Benisty},
  {Henning}, {van Boekel}, {Chapillon}, {Pinilla}, {Williams}, {Bertrang},
  {Facchini}, {Flock}, {Ginski}, {Juhasz}, {Klahr}, {Liu}, {M{\"u}ller},
  {P{\'e}rez}, {Pohl}, {Rosotti}, {Samland}, \&
  {Semenov}}]{2019A&A...625A.118K}
{Keppler}, M., {Teague}, R., {Bae}, J., {et~al.} 2019, \aap, 625, A118

\bibitem[{{Lee} {et~al.}(1996){Lee}, {Herbst}, {Pineau des Forets}, {Roueff},
  \& {Le Bourlot}}]{1996A&A...311..690L}
{Lee}, H.~H., {Herbst}, E., {Pineau des Forets}, G., {Roueff}, E., \& {Le
  Bourlot}, J. 1996, \aap, 311, 690

\bibitem[{{Lee} {et~al.}(1998){Lee}, {Roueff}, {Pineau des Forets},
  {Shalabiea}, {Terzieva}, \& {Herbst}}]{Lee_ea98}
{Lee}, H.-H., {Roueff}, E., {Pineau des Forets}, G., {et~al.} 1998, \aap, 334,
  1047

\bibitem[{{Lee} {et~al.}(2019){Lee}, {Lee}, {Baek}, {Aikawa}, {Cieza}, {Yoon},
  {Herczeg}, {Johnstone}, \& {Casassus}}]{Lee_ea19a}
{Lee}, J.-E., {Lee}, S., {Baek}, G., {et~al.} 2019, Nature Astronomy, 3, 314

\bibitem[{{Lin} \& {Papaloizou}(1993)}]{LP_93}
{Lin}, D.~N.~C. \& {Papaloizou}, J.~C.~B. 1993, in Protostars and Planets III,
  ed. E.~H. {Levy} \& J.~I. {Lunine}, 749

\bibitem[{{Long} {et~al.}(2019){Long}, {Herczeg}, {Harsono}, {Pinilla},
  {Tazzari}, {Manara}, {Pascucci}, {Cabrit}, {Nisini}, {Johnstone}, {Edwards},
  {Salyk}, {Menard}, {Lodato}, {Boehler}, {Mace}, {Liu}, {Mulders}, {Hendler},
  {Ragusa}, {Fischer}, {Banzatti}, {Rigliaco}, {van de Plas}, {Dipierro},
  {Gully-Santiago}, \& {Lopez-Valdivia}}]{Long_2019_Taurus_survey}
{Long}, F., {Herczeg}, G.~J., {Harsono}, D., {et~al.} 2019, \apj, 882, 49

\bibitem[{{Miotello} {et~al.}(2014){Miotello}, {Bruderer}, \& {van
  Dishoeck}}]{Miotello_ea14}
{Miotello}, A., {Bruderer}, S., \& {van Dishoeck}, E.~F. 2014, \aap, 572, A96

\bibitem[{{Miotello} {et~al.}(2018){Miotello}, {Facchini}, {van Dishoeck}, \&
  {Bruderer}}]{Miotello_ea18a}
{Miotello}, A., {Facchini}, S., {van Dishoeck}, E.~F., \& {Bruderer}, S. 2018,
  \aap, 619, A113

\bibitem[{{Miotello} {et~al.}(2019){Miotello}, {Facchini}, {van Dishoeck},
  {Cazzoletti}, {Testi}, {Williams}, {Ansdell}, {van Terwisga}, \& {van der
  Marel}}]{Miotello_ea19a}
{Miotello}, A., {Facchini}, S., {van Dishoeck}, E.~F., {et~al.} 2019, \aap,
  631, A69

\bibitem[{{Molyarova} {et~al.}(2018){Molyarova}, {Akimkin}, {Semenov},
  {{\'A}brah{\'a}m}, {Henning}, {K{\'o}sp{\'a}l}, {Vorobyov}, \&
  {Wiebe}}]{2018ApJ...866...46M}
{Molyarova}, T., {Akimkin}, V., {Semenov}, D., {et~al.} 2018, \apj, 866, 46

\bibitem[{{Molyarova} {et~al.}(2017){Molyarova}, {Akimkin}, {Semenov},
  {Henning}, {Vasyunin}, \& {Wiebe}}]{2017ApJ...849..130M}
{Molyarova}, T., {Akimkin}, V., {Semenov}, D., {et~al.} 2017, \apj, 849, 130

\bibitem[{{Padovani} {et~al.}(2018){Padovani}, {Ivlev}, {Galli}, \&
  {Caselli}}]{2018A&A...614A.111P}
{Padovani}, M., {Ivlev}, A.~V., {Galli}, D., \& {Caselli}, P. 2018, \aap, 614,
  A111

\bibitem[{{Pagani} {et~al.}(2009){Pagani}, {Vastel}, {Hugo}, {Kokoouline},
  {Greene}, {Bacmann}, {Bayet}, {Ceccarelli}, {Peng}, \&
  {Schlemmer}}]{2009A&A...494..623P}
{Pagani}, L., {Vastel}, C., {Hugo}, E., {et~al.} 2009, \aap, 494, 623

\bibitem[{{Parfenov} {et~al.}(2016){Parfenov}, {Semenov}, {Sobolev}, \&
  {Gray}}]{2016MNRAS.460.2648P}
{Parfenov}, S.~Y., {Semenov}, D.~A., {Sobolev}, A.~M., \& {Gray}, M.~D. 2016,
  \mnras, 460, 2648

\bibitem[{{Pavlyuchenkov} {et~al.}(2007){Pavlyuchenkov}, {Semenov}, {Henning},
  {Guilloteau}, {Pi{\'e}tu}, {Launhardt}, \& {Dutrey}}]{2007ApJ...669.1262P}
{Pavlyuchenkov}, Y., {Semenov}, D., {Henning}, T., {et~al.} 2007, \apj, 669,
  1262

\bibitem[{{Qi} {et~al.}(2013){Qi}, {{\"O}berg}, \& {Wilner}}]{Qi_ea13a}
{Qi}, C., {{\"O}berg}, K.~I., \& {Wilner}, D.~J. 2013, \apj, 765, 34

\bibitem[{Rab {et~al.}(2017)Rab, G{\"{u}}del, Padovani, Kamp, Thi, Woitke, \&
  Aresu}]{Rab2017}
Rab, C., G{\"{u}}del, M., Padovani, M., {et~al.} 2017, A{\&}A, 603, 96

\bibitem[{{Rab} {et~al.}(2018){Rab}, {G{\"u}del}, {Woitke}, {Kamp}, {Thi},
  {Min}, {Aresu}, \& {Meijerink}}]{Rab_ea_2018}
{Rab}, C., {G{\"u}del}, M., {Woitke}, P., {et~al.} 2018, \aap, 609, A91

\bibitem[{{Roberts} {et~al.}(2004){Roberts}, {Herbst}, \&
  {Millar}}]{2004A&A...424..905R}
{Roberts}, H., {Herbst}, E., \& {Millar}, T.~J. 2004, \aap, 424, 905

\bibitem[{{Roberts} \& {Millar}(2000)}]{2000A&A...361..388R}
{Roberts}, H. \& {Millar}, T.~J. 2000, \aap, 361, 388

\bibitem[{{R{\"o}llig} {et~al.}(2007){R{\"o}llig}, {Abel}, {Bell}, {Bensch},
  {Black}, {Ferland}, {Jonkheid}, {Kamp}, {Kaufman}, {Le Bourlot}, {Le Petit},
  {Meijerink}, {Morata}, {Ossenkopf}, {Roueff}, {Shaw}, {Spaans}, {Sternberg},
  {Stutzki}, {Thi}, {van Dishoeck}, {van Hoof}, {Viti}, \&
  {Wolfire}}]{Roellig_gas_thermal_balance}
{R{\"o}llig}, M., {Abel}, N.~P., {Bell}, T., {et~al.} 2007, \aap, 467, 187

\bibitem[{{Roueff} {et~al.}(2005){Roueff}, {Lis}, {van der Tak}, {Gerin}, \&
  {Goldsmith}}]{2005A&A...438..585R}
{Roueff}, E., {Lis}, D.~C., {van der Tak}, F.~F.~S., {Gerin}, M., \&
  {Goldsmith}, P.~F. 2005, \aap, 438, 585

\bibitem[{{Salinas} {et~al.}(2018){Salinas}, {Hogerheijde}, {Murillo},
  {Mathews}, {Qi}, {Williams}, \& {Wilner}}]{Salinas_2018_DCOp}
{Salinas}, V.~N., {Hogerheijde}, M.~R., {Murillo}, N.~M., {et~al.} 2018, \aap,
  616, A45

\bibitem[{{Sch{\"o}ier} {et~al.}(2005){Sch{\"o}ier}, {van der Tak}, {van
  Dishoeck}, \& {Black}}]{lamda}
{Sch{\"o}ier}, F.~L., {van der Tak}, F.~F.~S., {van Dishoeck}, E.~F., \&
  {Black}, J.~H. 2005, \aap, 432, 369

\bibitem[{{Semenov} {et~al.}(2010){Semenov}, {Hersant}, {Wakelam}, {Dutrey},
  {Chapillon}, {Guilloteau}, {Henning}, {Launhardt}, {Pi{\'e}tu}, \&
  {Schreyer}}]{Semenov_ea_2010}
{Semenov}, D., {Hersant}, F., {Wakelam}, V., {et~al.} 2010, \aap, 522, A42

\bibitem[{{Semenov} \& {Wiebe}(2011)}]{Semenov_Wiebe11}
{Semenov}, D. \& {Wiebe}, D. 2011, \apj, 196, 25

\bibitem[{{Semenov} {et~al.}(2004){Semenov}, {Wiebe}, \&
  {Henning}}]{Semenov_ea04}
{Semenov}, D., {Wiebe}, D., \& {Henning}, T. 2004, \aap, 417, 93

\bibitem[{{Sipil{\"a}} {et~al.}(2013){Sipil{\"a}}, {Caselli}, \&
  {Harju}}]{2013A&A...554A..92S}
{Sipil{\"a}}, O., {Caselli}, P., \& {Harju}, J. 2013, \aap, 554, A92

\bibitem[{{Teague} {et~al.}(2018){Teague}, {Bae}, {Birnstiel}, \&
  {Bergin}}]{2018ApJ...868..113T}
{Teague}, R., {Bae}, J., {Birnstiel}, T., \& {Bergin}, E.~A. 2018, \apj, 868,
  113

\bibitem[{{Teague} \& {Loomis}(2020)}]{Teague_Loomis20}
{Teague}, R. \& {Loomis}, R. 2020, arXiv e-prints, arXiv:2007.11906

\bibitem[{{Teague} {et~al.}(2017){Teague}, {Semenov}, {Gorti}, {Guilloteau},
  {Henning}, {Birnstiel}, {Dutrey}, {van Boekel}, \&
  {Chapillon}}]{Teague_ea_2017}
{Teague}, R., {Semenov}, D., {Gorti}, U., {et~al.} 2017, \apj, 835, 228

\bibitem[{{Teague} {et~al.}(2015){Teague}, {Semenov}, {Guilloteau}, {Henning},
  {Dutrey}, {Wakelam}, {Chapillon}, \& {Pietu}}]{Teague_ea_2015}
{Teague}, R., {Semenov}, D., {Guilloteau}, S., {et~al.} 2015, \aap, 574, A137

\bibitem[{{van Boekel} {et~al.}(2017){van Boekel}, {Henning}, {Menu}, {de
  Boer}, {Langlois}, {M{\"u}ller}, {Avenhaus}, {Boccaletti}, {Schmid},
  {Thalmann}, {Benisty}, {Dominik}, {Ginski}, {Girard}, {Gisler}, {Lobo Gomes},
  {Menard}, {Min}, {Pavlov}, {Pohl}, {Quanz}, {Rabou}, {Roelfsema}, {Sauvage},
  {Teague}, {Wildi}, \& {Zurlo}}]{2017ApJ...837..132V}
{van Boekel}, R., {Henning}, T., {Menu}, J., {et~al.} 2017, \apj, 837, 132

\bibitem[{{van Terwisga} {et~al.}(2019){van Terwisga}, {van Dishoeck},
  {Cazzoletti}, {Facchini}, {Trapman}, {Williams}, {Manara}, {Miotello}, {van
  der Marel}, {Ansdell}, {Hogerheijde}, {Tazzari}, \&
  {Testi}}]{vTerwisga_ea19a}
{van Terwisga}, S.~E., {van Dishoeck}, E.~F., {Cazzoletti}, P., {et~al.} 2019,
  \aap, 623, A150

\bibitem[{{Vasyunin} \& {Herbst}(2013)}]{2013ApJ...769...34V}
{Vasyunin}, A.~I. \& {Herbst}, E. 2013, \apj, 769, 34

\bibitem[{{Vasyunin} {et~al.}(2008){Vasyunin}, {Semenov}, {Henning}, {Wakelam},
  {Herbst}, \& {Sobolev}}]{Vasyunin_ea08}
{Vasyunin}, A.~I., {Semenov}, D., {Henning}, T., {et~al.} 2008, \apj, 672, 629

\bibitem[{{Visser} {et~al.}(2009){Visser}, {van Dishoeck}, \&
  {Black}}]{Visser_ea09b}
{Visser}, R., {van Dishoeck}, E.~F., \& {Black}, J.~H. 2009, Astron.~Astrophys,
  503, 323

\bibitem[{{Wakelam} {et~al.}(2012){Wakelam}, {Herbst}, {Loison}, {Smith},
  {Chandrasekaran}, {Pavone}, {Adams}, {Bacchus-Montabonel}, {Bergeat},
  {B{\'e}roff}, {Bierbaum}, {Chabot}, {Dalgarno}, {van Dishoeck}, {Faure},
  {Geppert}, {Gerlich}, {Galli}, {H{\'e}brard}, {Hersant}, {Hickson},
  {Honvault}, {Klippenstein}, {Le Picard}, {Nyman}, {Pernot}, {Schlemmer},
  {Selsis}, {Sims}, {Talbi}, {Tennyson}, {Troe}, {Wester}, \&
  {Wiesenfeld}}]{KIDA}
{Wakelam}, V., {Herbst}, E., {Loison}, J.-C., {et~al.} 2012,
  Astrophys.~J.,~Suppl.~Ser, 199, 21

\bibitem[{{Walmsley} {et~al.}(2004){Walmsley}, {Flower}, \& {Pineau des
  For{\^e}ts}}]{2004A&A...418.1035W}
{Walmsley}, C.~M., {Flower}, D.~R., \& {Pineau des For{\^e}ts}, G. 2004, \aap,
  418, 1035

\bibitem[{{Walsh} {et~al.}(2015){Walsh}, {Nomura}, \& {van
  Dishoeck}}]{2015A&A...582A..88W}
{Walsh}, C., {Nomura}, H., \& {van Dishoeck}, E. 2015, \aap, 582, A88

\bibitem[{{Wiebe} {et~al.}(2003){Wiebe}, {Semenov}, \& {Henning}}]{Wiebe_ea03a}
{Wiebe}, D., {Semenov}, D., \& {Henning}, T. 2003, \aap, 399, 197

\bibitem[{{Wilson} \& {Rood}(1994)}]{Wilson_Rood94}
{Wilson}, T.~L. \& {Rood}, R. 1994, \araa, 32, 191

\bibitem[{{Woitke} {et~al.}(2009){Woitke}, {Kamp}, \&
  {Thi}}]{Woitke_Kamp_Thi_2009_ProDiMo}
{Woitke}, P., {Kamp}, I., \& {Thi}, W.~F. 2009, \aap, 501, 383

\bibitem[{{Yorke} \& {Bodenheimer}(2008)}]{Yorke_Bodenheimer_2008_track}
{Yorke}, H.~W. \& {Bodenheimer}, P. 2008, in Astronomical Society of the
  Pacific Conference Series, Vol. 387, Massive Star Formation: Observations
  Confront Theory, ed. H.~{Beuther}, H.~{Linz}, \& T.~{Henning}, 189

\end{thebibliography}
\bibliographystyle{aa}

\end{document}